\documentclass[aps,prd,twocolumn,floatfix,noshowpacs,tightenlines,noshowkeys,superscriptaddress,amsmath,amssymb,nofootinbib]{revtex4-2}

\usepackage{txfonts}

\usepackage{graphicx}
\usepackage{xcolor}
\usepackage{dcolumn}
\usepackage{hyperref}
\usepackage{amssymb}
\usepackage{xspace}
\usepackage{amsmath}
\usepackage{soul}
\usepackage[mathlines]{lineno}
\usepackage{aas_macros}
\usepackage{orcid}
\usepackage{enumitem}
\usepackage{booktabs,dcolumn,makecell}

\newcommand{\Msun}{\ensuremath{M_{ \odot }}}

\newcommand{\ppop}{\ensuremath{p_{\rm pop}}\xspace}

\newcommand{\hu}{\xspace \ensuremath{{\rm km \, s^{-1} \, Mpc^{-1}}\xspace}}

\newcommand{\sg}{\ensuremath{\hat{\alpha}_0} \xspace}

\newcommand{\de}{{\rm d}}

\newcommand{\icarogw}{\textsc{icarogw}\xspace} 

\newcommand{\Nobs}{\ensuremath{N_{\rm obs}}\xspace}
\newcommand{\Vd}{\ensuremath{\vec{d}}\xspace}
\newcommand{\Vdi}{\ensuremath{\vec{d}_{\rm i}}\xspace}
\newcommand{\Vlambda}{\ensuremath{\vec{\lambda}}\xspace}
\newcommand{\Vtheta}{\ensuremath{\vec{\theta}}\xspace}
\newcommand{\like}{\ensuremath{\mathcal{L}}\xspace}
\newcommand{\td}{\ensuremath{t_{\rm d}}\xspace}
\newcommand{\ts}{\ensuremath{t_{\rm s}}\xspace}
\newcommand{\Dl}{\ensuremath{D_{\rm L}}\xspace}
\newcommand{\Dtd}{\ensuremath{\Delta t_{\rm d}}\xspace}
\newcommand{\Dts}{\ensuremath{\Delta t_{\rm s}}\xspace}
\newcommand{\z}{\ensuremath{z}\xspace}

\newcommand{\pdet}{\ensuremath{p_{\rm det}}\xspace}
\newcommand{\vl}{\ensuremath{|}\xspace}
\newcommand{\HO}{\ensuremath{H_{0}}\xspace}
\newcommand{\Om}{\ensuremath{\Omega_{m}}\xspace}
\newcommand{\Tl}{\ensuremath{ \text{T}_l}\xspace}
\newcommand{\aO}{\ensuremath{\hat{\alpha}_0}\xspace}
\newcommand{\vgw}{\ensuremath{v_{\rm GW}}\xspace}

\newcommand{\ls}{\ensuremath{c}\xspace}
\newcommand{\LCDM}{\ensuremath{\Lambda\text{CDM}}\xspace}

\begin{document}

\title{Measuring the speed of gravity and the cosmic expansion with time delays between gravity and light from binary neutron stars}

\author{Leonardo Iampieri \orcidlink{0009-0004-1161-2990}}
\email{leonardo.iampieri@uniroma1.it}
\affiliation{Dipartimento di Fisica, Università di Roma ``Sapienza,'' Piazzale A. Moro 5, I-00185, Roma, Italy}
\affiliation{INFN, Sezione di Roma, I-00185 Roma, Italy}

\author{Simone Mastrogiovanni \orcidlink{0000-0003-1606-4183}}
\affiliation{INFN, Sezione di Roma, I-00185 Roma, Italy}

\author{Francesco Pannarale \orcidlink{0000-0002-7537-3210}}
\affiliation{Dipartimento di Fisica, Università di Roma ``Sapienza,'' Piazzale A. Moro 5, I-00185, Roma, Italy}
\affiliation{INFN, Sezione di Roma, I-00185 Roma, Italy}

\keywords{}

\date{\today}

\begin{abstract}
The first observation of a gravitational wave (GW) and a short gamma-ray burst (sGRB) emitted by the same binary neutron star (BNS) merger officially opened the field of GW multimessenger astronomy. In this paper, we define and address \textit{lagging sirens}, a new class of multimessenger BNSs for which associated GWs and sGRBs are observed without the identification of their host galaxy. We propose a new methodology to use the observed time delay of these sources to constrain the speed of gravity that is, the propagation speed of gravitational waves, the Hubble constant and the prompt time delay distribution between GWs and sGRBs, even though a direct redshift estimation from the host galaxy is unavailable. Our method exploits the intrinsic relation between GWs and sGRBs observed and prompt time delays to obtain a statistical redshift measure for the cosmological sources. We show that this technique can be used to infer the Hubble constant at the $10\%$~level of precision with future-generation GW detectors such as the Einstein Telescope and only 100 observations of this kind. The novel procedure that we propose has systematics that differ completely from the ones of previous GW methods for cosmology. Additionally, we demonstrate for the first time that the speed of gravity and the distribution of the prompt time delays between GWs and sGRBs can be inferred conjointly with less than 10 sources even with current GW detector sensitivities.
\end{abstract}

\maketitle


\section{Introduction}

 Gravitational waves (GWs) from compact binary coalescences (CBCs) can be used to probe the expansion of the Universe. Given the current open tensions in cosmology \cite{2022NatAs...6.1353D}, it is crucial to measure the Universe expansion parameters, particularly the Hubble constant $H_0$, with independent classes of sources. CBCs detected via GWs are the only cosmological sources for which a direct measurement of the luminosity distance without the need for an astrophysical calibration is possible \cite{Schutz:1986gp, 2005ApJ...629...15H}. For this reason, GW sources are referred to as ``standard sirens.'' Unfortunately, GWs do not directly provide the source redshift which is the second ingredient required to measure the cosmic expansion parameters. Current scientific literature proposes and uses several methodologies to assign redshifts to GW events and infer the cosmic expansion parameters.

In terms of analysis, the most trivial scenario is the one of \emph{bright sirens}, where an electromagnetic (EM) counterpart is observed conjointly with the GW emission. This is the case of the binary neutron star (BNS) merger GW170817 \cite{Abbott_2017, 2017ApJ...848L..12A}. For these cases, it could be possible to identify the host galaxy and obtain a spectroscopic estimate of the galaxy redshift. The observation of GW170817 and its EM counterpart was exploited to obtain $H_0=70^{+19}_{-8} \hu$ \cite{LIGOScientific:2017adf, 2019PhRvX...9a1001A}. The error budget on the $H_0$ measure is entirely dominated by the luminosity distance error from the GW detection \cite{2019PhRvD.100h3514C}, as the redshift is accurately measured.

So far, GW170817 is the only GW event with an unambiguously associated EM counterpart. Out of almost 100 GW detections achieved to date \cite{LIGOScientific:2020ibl, LIGOScientific:2021usb, KAGRA:2021duu}, most are due to binary black hole mergers observed without an EM counterpart (\emph{dark sirens}). Among the several techniques proposed for dark sirens cosmology is the source-frame mass method \cite{Taylor:2011fs, Wysocki:2018mpo, Farr:2019twy, Mastrogiovanni:2021wsd, Mancarella:2021ecn, Mukherjee:2021rtw, Gray:2021qfw, Gray:2021sew, Gair:2022zsa, Leyde:2022orh, Ezquiaga:2022zkx,2023arXiv231205302B,2023MNRAS.523.4539K}. This method consists of exploiting the difference between the detector mass $m_d$ and source mass $m_s$ of CBCs, which are related by a redshift factor $m_d=(1+z)m_s$. Thus, by knowing the source-frame mass spectrum, it is possible to infer statistically the source redshift. Since the source mass spectrum is not known accurately, current studies conjointly fit flexible phenomenological models of the mass spectrum along with cosmological parameters that can in some limits approximate the true binary distributions \cite{2023arXiv231211627P}. The source frame mass method was used with the latest GW detections to obtain a value of $H_0=68^{+12}_{-7} \hu$ \cite{LIGOScientific:2021aug}. The low precision in the inferred value of $H_0$ is due to the broadness of the source-frame mass spectrum, which prevents us from obtaining precise redshift estimations for the GW event sources used to measure $H_0$.

Following the same idea of the source-frame mass method, this paper proposes a new method for GW cosmology: the \textit{prompt time delay} method. The methodology consists of obtaining an implicit source redshift information for GW sources with observed short gamma-ray bursts (sGRBs) but no identified host galaxy. We refer to this type of source as \textit{lagging sirens}. Currently, there are no GW detections of this kind. However, we certainly expect them for future GW detectors such as Einstein Telescope (ET) \cite{2010CQGra..27s4002P, Maggiore:2019uih}. ET will detect BNSs up to redshift 2-3 \cite{2023JCAP...07..068B}. Although the farthest GRBs, and their afterglows, could be detected even at these redshifts \cite{2009A&A...507L..45A}, it might not be trivial to identify and obtain a spectroscopic measurement of its host galaxy. 

The paper is organized as follows. In Sec.~\ref{sec:2}, we introduce the relations between prompt and observed GW-sGRB time delays considering also possible deviations between the speed of light and gravity. We further introduce the statistical framework used to infer the Hubble constant $H_0$, speed of gravity $\hat{\alpha}_0$ and prompt time delay distribution. In Sec.~\ref{sec:3} we simulate catalogs of GW and sGRB observations considering two scenarios for the sensitivity of GW detectors, namely, the sensitivity expected to be achieved in 2027-2030 with the fifth LIGO-Virgo-KAGRA Observing Run (O5) \cite{2015, Acernese_2014, 10.1093/ptep/ptaa125, KAGRA:2013rdx} and an ET-like sensitivity expected to be achieved in 2035+. In Sec.~\ref{sec:4} we present the prospects of measuring $H_0$, \sg and the prompt time delay distribution. Finally, in Sec.~\ref{sec:5} we draw our conclusions.

\section{Cosmological and statistical models}
\label{sec:2}

This section provides the background knowledge required to build the prompt time delay method. Specifically, in Sec.~\ref{sec:2a}, we introduce the relations between cosmological expansion, redshift and prompt time delay distributions, while in Sec.~\ref{sec:2b} we review the Bayesian statistical framework used in this work.

\subsection{Theoretical background}
\label{sec:2a}

We model the cosmic expansion with a flat \LCDM model described by a Hubble constant $H_0$, dark matter energy density $\Omega_m$ and dark energy density $\Omega_\Lambda=1-\Omega_m$. In this model, the luminosity distance $\Dl$ and redshift $z$ of cosmological sources are related by
\begin{equation}
\label{eq:Dl}
    \Dl(z)=\frac{\ls(1+z)}{H_0} \int_0^z \frac{\de z'}{\sqrt{\Omega_m(1+z')^3+(1-\Omega_m)}}\,,
\end{equation}
where $c$ is the speed of light.

We define the prompt time delay \Dts as the elapsed time between the GW luminosity peak at the CBC merger and the emission of the sGRB. A positive \Dts indicates that the sGRB is emitted after the CBC merger, while a negative value indicates that it is emitted prior to the merger. The observed time delay between GW and sGRB at the merger is \cite{2020PhRvD.102d4009M}
\begin{equation}
\label{eq:time_delay}
\Dtd = (1 + z) \Dts + \frac{\aO} {2} \Tl\,,
\end{equation}
where \Tl is the lookback time
\begin{equation}
\label{eq:lookback}
\Tl = \frac{1}{\HO}\int_0^{z} \de z' \frac{1}{\sqrt{\Omega_m(1+z')^3+(1-\Omega_m)}}\,,
\end{equation}
and \aO is a scalar factor proportional to the difference between the GW's speed \vgw and the speed of light relative to the speed of light itself, namely
\begin{equation}\label{eq:alpha_0}
\aO = 2 \frac{\vgw- c}{c}\,.
\end{equation}

In writing Eq.~\eqref{eq:time_delay}, we implicitly assumed that the speed of gravity does not depend on the GW frequency and that its deviation from the speed of light is small \cite{2020PhRvD.102d4009M}. Moreover, we assumed that the speed of gravity does not change during cosmic time, although this possibility is not excluded in principle for some theories of gravity \cite{Romano:2023bge,2024PhLB..85138572R} or agnostic tests on the speed of gravity \cite{PhysRevD.108.124017}.

\begin{figure}
    \centering
    \includegraphics[scale=0.25]{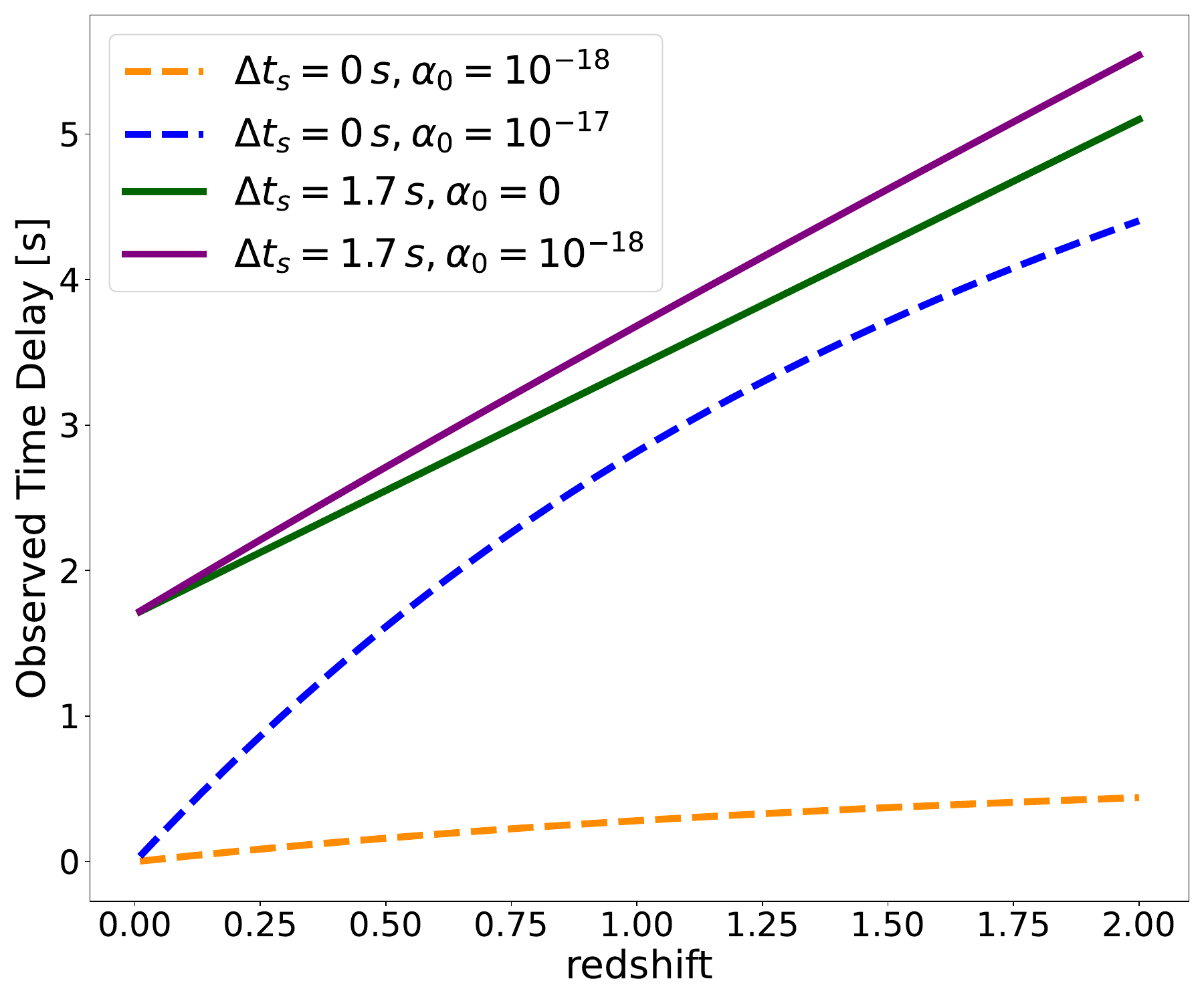}
    \caption{Observed time delays (vertical axis) as a function of redshift (horizontal axis) for different values of the speed of gravity parameter and prompt time delay. The different lines indicate various choices for the prompt time delay and speed of gravity.}
    \label{fig:observed_time_delay_comparison}
\end{figure}

Equation \eqref{eq:time_delay} can be used to predict the observed time delays between the GW and sGRB signals from the same event. For instance, if the GW and sGRB were prompted together ($\Dts=0$ s), and GWs were slightly slower than the light ($\sg<0$), then the GW would be detected after the arrival of the sGRB ($\Dtd<0$ s). If instead the sGRB were prompted after the GW ($\Dts>0$ s) and the speed of gravity were equal to $c$ ($\sg=0$), then the GW would surely be detected prior to the sGRB ($\Dtd>0$ s).
Figure \ref{fig:observed_time_delay_comparison} shows the observed time delays as a function of redshift for various values of the speed of gravity and prompt time delay. The larger the redshift, the longer the observed time delay between the GW and the sGRB is. From Fig.~\ref{fig:observed_time_delay_comparison}, we can already see that if the observed time delay were accurately measured\footnote{This is a reasonable assumption also supported by the measure of GW170817 with its sGRBs for which an observed time delay of $\Dtd = 1.74 \pm 0.05$ s was observed \cite{Abbott_2017}.} and the prompt time delay and speed of gravity were known, one could get implicit redshift information for the GW-sGRB sources. Unfortunately, while we can argue that the speed of gravity must be equal to the speed of light if we assume classical general relativity, the distribution of prompt time delays between GWs and sGRBs is not a constant and it is currently unknown. For this reason, we introduce a statistical framework able to infer it.

\subsection{Statistical background}
\label{sec:2b}

Let us assume that we have a set of common population parameters \Vlambda, such as the Hubble constant or the propagation speed of GWs, that we want to infer from \Nobs GW-sGRB joint observations collected from the data \{$\Vd_1, \ldots, \Vd_{\Nobs}$\}. We further assume that each astrophysical source is described by a set of parameters \Vtheta, the distributions of which are described by a CBC rate model
\begin{equation}
    \frac{\de N}{\de\td \de\Vtheta}(\Vtheta\vl\Vlambda)\,,
\end{equation}
where $\td$ is the detector time. Our aim is to infer the parameters \Vlambda by conjointly fitting the rate distribution of the parameters \Vtheta.
The hierarchical likelihood of obtaining these detections is \citep{Mandel_2019}
\begin{equation}\label{eq:Mandel}
\like(\{\Vdi\}\vl\Vlambda) = \prod_{i=1}^{\Nobs} \frac{\int \like(\Vdi\vl\Vtheta,\Vlambda) \frac{\de N}{\de\td \de\Vtheta}(\Vtheta\vl\Vlambda)  \; \de\Vtheta \; \de\td}{\int \pdet(\Vtheta,\Vlambda) \frac{\de N}{\de\td \de\Vtheta}(\Vtheta\vl\Vlambda)  \; \de\Vtheta \; \de\td}\,,
\end{equation}
where \like(\Vdi\vl\Vtheta,\Vlambda) is the measurement likelihood, which quantifies how well we can measure \Vtheta, conditioning on \Vlambda, from a single event detection \Vdi, while \pdet(\Vtheta,\Vlambda) is a detection probability used to model the finite sensitivity of GW and sGRB detectors.

For this study, the relevant \Vtheta parameters are the luminosity distance \Dl and the observed GW-sGRB time delay \Dtd. Thus, the CBC rate model written in terms of these two quantities is
\begin{equation}\label{eq:rate}
\frac{\de N}{\de\td \de\Vtheta} \equiv \frac{\de N}{\de\td \de\Dl \de\Dtd}\,.
\end{equation}
Since we want to exploit the prompt time delay distribution \Dts to obtain an implicit redshift estimation $z$, we need to rewrite the CBC rate in terms of these quantities. This can be done by performing a change of coordinates from luminosity distance and observed time delay to redshift and prompt time delay, namely
\begin{equation}\label{eq:det_rate}
\frac{\de N}{\de\td \de\Dl \de\Dtd} \equiv \frac{\de N}{\de\ts \de\z \de\Dts} \frac{1}{1 + z} \frac{1}{\text{det} J_{D \rightarrow S}}\,,
\end{equation}
where the \((1 + \z)^{-1}\) factor arises from the transformation between source frame $t_s$ and detector frame time $t_d$, $\frac{\de N}{\de\ts \de\z \de\Dts}$ is the source frame rate per redshift \z and per prompt time delay \Dts, while \( \text{det} J_{D \rightarrow S}\) is the determinant of the Jacobian associated to the change of variables between the detector and source frames (including also GW-sGRB time delays), that is,
\begin{equation}\label{eq:jacobian}
\text{det} J_{D \rightarrow S} = \left|\frac{\partial \Dl}{\partial \z} \frac{\partial \Dtd}{\partial \Dts}\right| = \left|\frac{\partial D_L}{\partial z} (1 + z)\right|\,,
\end{equation}
where
\begin{equation}
\frac{\partial \Dl}{\partial \z} = \frac{\Dl(\z)}{1 + \z} + \frac{c(1 + \z)}{\HO} \frac{1}{\sqrt{\Omega_m(1+z)^3+(1-\Omega_m)}}\,.
\end{equation}
We additionally parameterize the source frame rate per redshift \z and per prompt time delay as 
\begin{equation}\label{eq:source_rate}
\begin{split}
\frac{\de N}{\de \ts \de\z \de\Dts} &= \frac{\de N}{\de\ts \de V_c \de\Dts} \frac{\de V_c}{\de\z} = \\ &= R_0  \psi(\z;\Vlambda) \ppop(\Dts\vl\Vlambda) \frac{\de V_c}{\de\z},
\end{split}
\end{equation}
where $R_0$ is the BNS merger rate per comoving volume at redshift $z=0$, $\psi(z;\Vlambda)$ is a phenomenological parametrization for the BNS merger rate as a function of redshift such that $\psi(z=0;\vec{\lambda})=1$, \ppop is a probability density function describing the prompt time delay distribution, and
\begin{equation}
 \frac{\de V_c}{\de z } = 4 \pi \left[\frac{c}{H_0}\right]^3 \left[\int_0^z \frac{\de z'}{E(z')}\right]^2 
\end{equation}
is the differential of the comoving volume.\\
The overall detector rate model written in terms of source rate quantities can be rewritten as
\begin{equation}\label{eq:full_rate}
\frac{\de N}{\de\td \de\Dl \de\Dtd} =  R_0  \psi(\z;\Vlambda) \ppop(\Dts\vl\Vlambda) \frac{\de V_c}{\de\z} \frac{1}{(1 + \z)^2\left|\frac{\partial \Dl}{\partial \z}\right|}\,.
\end{equation}

By substituting Eq.~\eqref{eq:full_rate} in Eq.~\eqref{eq:Mandel}, we find the full expression of the hierarchical likelihood
\begin{equation}\label{eq:full_likelihood}
\like(\{\Vdi\}\vl\Vlambda) \propto \prod_{i=1}^{\Nobs} \frac{\int \like(\Vdi\vl\Dl,\Dtd,\Vlambda)  \frac{\de V_c}{\de\z} \frac{\psi(\z;\Vlambda) \ppop(\Dts\vl\Vlambda) }{(1 + \z)^2\left|\frac{\partial \Dl}{\partial \z}\right|}  \; \de\Dl \de\Dtd }{\int \pdet(\Dl,\Dtd,\Vlambda) \frac{\de V_c}{\de\z} \frac{\psi(\z;\Vlambda) \ppop(\Dts\vl\Vlambda) }{(1 + \z)^2\left|\frac{\partial \Dl}{\partial \z}\right|}  \; \de\Dl \de\Dtd}\,.
\end{equation}

To evaluate the hierarchical likelihood, we use \icarogw \cite{mastrogiovanni2023icarogw}, a software able to calculate an approximate value of Eq.~\eqref{eq:full_likelihood} starting from a set of posterior samples on luminosity distances and on observed time delays, and detectable injections used to evaluate the selection bias.

\subsection{Study case with GW170817 and GRB 170817A}

To display an application of this method, we recalculate the constraints on the speed of gravity obtained in \cite{Abbott_2017} from the joint observation of GW170817 and GRB 170817A. We consider two cases: the first in which we only consider the luminosity distance estimated from the GW signal and the observed time delay between GW170817 and GRB 170817A and the second in which we also include the redshift information from the event host galaxy. To include the redshift information from the host galaxy, the hierarchical likelihood in Eq.~\eqref{eq:full_likelihood} can be modified as follows
\begin{widetext}
\begin{equation}\label{eq:full_likelihood_2}
\like(\{\Vdi\}\vl\Vlambda) \propto \prod_{i=1}^{\Nobs} \frac{\int \like(\Vdi\vl\Dl,\Dtd,\Vlambda) \like(\vec{d}_i^{\rm EM}|z,\sigma_z)  \frac{\de V_c}{\de\z} \frac{\psi(\z;\Vlambda) \ppop(\Dts\vl\Vlambda) }{(1 + \z)^2\left|\frac{\partial \Dl}{\partial \z}\right|}  \; \de\Dl \de\Dtd }{\int \pdet(\Dl,\Dtd,\Vlambda) \frac{\de V_c}{\de\z} \frac{\psi(\z;\Vlambda) \ppop(\Dts\vl\Vlambda) }{(1 + \z)^2\left|\frac{\partial \Dl}{\partial \z}\right|}  \; \de\Dl \de\Dtd}\,.
\end{equation}
\end{widetext}
where $\like(\vec{d}_i^{\rm EM}|z,\sigma_z)$ encapsulates the measure of the redshift from the source host galaxy. Following \cite{LIGOScientific:2017adf}, we assume this is a Gaussian distribution with mean $z=0.0100$ and $\sigma=0.0005$. Note also that in Eq.~\eqref{eq:full_likelihood_2}, the detection probability at the denominator should also contain the detection probability of the EM counterpart. However, we neglect this component as during the O2 run, the detection range of electromagnetic counterparts was significantly higher than the one of the GW sources. To estimate the source luminosity distance, and hence the redshift, we use the low-spin parameter estimation samples released\footnote{\url{https://dcc.ligo.org/LIGO-P1800061/public}} with \cite{2019PhRvX...9a1001A}. The estimated value of the luminosity distance from the posterior samples is $D_{\rm L}=41^{+6}_{-12}$ Mpc \cite{2019PhRvX...9a1001A}. We use a flat $\Lambda$CDM model with cosmological parameters from \cite{2020} ($H_0 = 67.66 \, \hu$, $\Om = 0.310 $) to convert between luminosity distance and redshift. We also assume that the time delay at the detector between GW170817 and GRB 170817A is $\Delta t_d = 1.74^{+0.05}_{-0.05}$ s, measured with a Gaussian posterior. More importantly, we fix the distribution of source prompt time delays to be uniform between [0,10] s, meaning that the GRB can be emitted up to 10 seconds after the merger. This choice is made following \cite{Abbott_2017}, where to derive the constraints on the speed of gravity it was assumed that the GRB could be emitted at most 10 seconds after the merger.

\begin{figure}[!tb]
    \centering
    \includegraphics[width=0.5\textwidth]{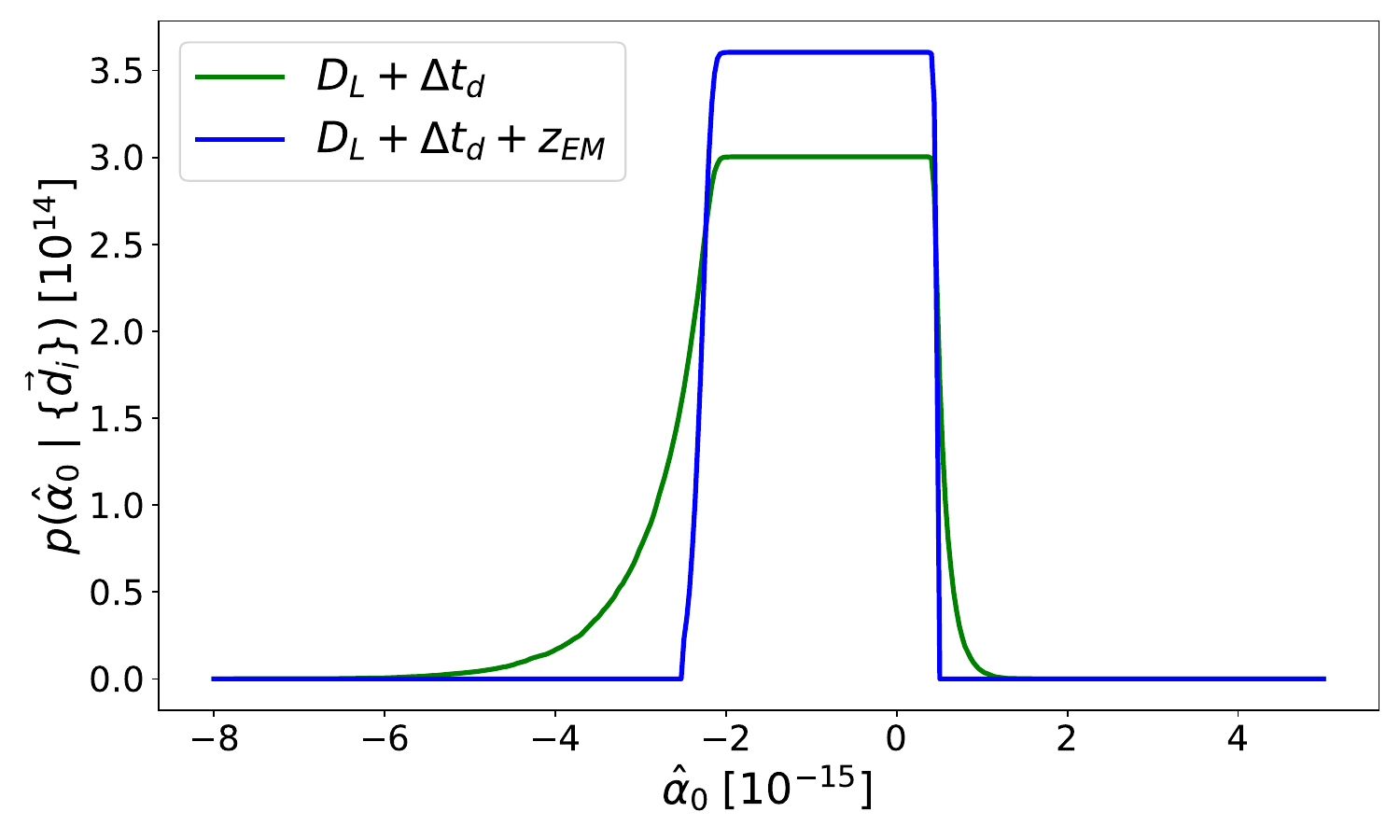}
    \caption{Posterior distributions for the speed of gravity from the joint observation of GW170817 and GRB 170817A. The green line represents the posterior distribution using only the measurement of luminosity Distance (\Dl) and Time Delay (\Dtd). The blue line represents the posterior distribution when including the redshift information from the event host galaxy.}
    \label{fig:GW170817}
\end{figure}

Figure \ref{fig:GW170817} shows the posterior distribution of the fractional difference between the speed of gravity and the speed of light. The boundaries of the posterior can be constrained as a proxy for the constraints on the speed of gravity. When just using GW170817 and GRB 170817A, we obtain
\begin{equation}
-3 \times 10^{-15} \lesssim \frac{\vgw- c}{c}\ \lesssim +7 \times 10^{-16},
\end{equation}
which matches the constraints obtained in \cite{Abbott_2017} by combining only GW and Fermi-INTEGRAL data. Notably, in \cite{Abbott_2017}, the authors excluded inputs from external analyses, such as the identification of NGC 4993 via EM observations at lower wavelengths, and avoided assumptions about cosmology required to convert the luminosity distance into redshift. Instead, they assumed a conservative (in terms of speed of gravity constraints) lower limit of 26 Mpc for the luminosity distance. When we also include the information on the redshift of NGC 4993, we obtain 
\begin{equation}
-1 \times 10^{-15} \lesssim \frac{\vgw- c}{c}\ \lesssim +3 \times 10^{-16}.
\end{equation}
We can see that the posterior generated including the redshift estimation of NGC 4993 has shorter tails than the one generated without redshift. 
This is due to the fact that, the redshift of GW170817 is more precise than the luminosity distance inferred from the GW. As a consequence, when we fix the cosmology, the luminosity distance of the source is almost entirely given by the galaxy host redshift, with a negligible uncertainty on its value.

\section{Mock Catalogs of GW and sGRB detections}
\label{sec:3}

We simulated mock catalogs of GW and sGRB detections to forecast the potential of the method we propose to constrain prompt time delay distributions, speed of gravity and $H_0$. In Sec.~\ref{sec:3a} we describe how we simulate joint detections of GW and sGRB, while in Sec.~\ref{sec:3b} we provide a general overview of the observational properties in terms of luminosity distance and observed time delay of the GW-sGRB detections.

\subsection{Catalogs generation}
\label{sec:3a}

\begin{figure}
    \centering
    \includegraphics[width=0.275\textwidth]{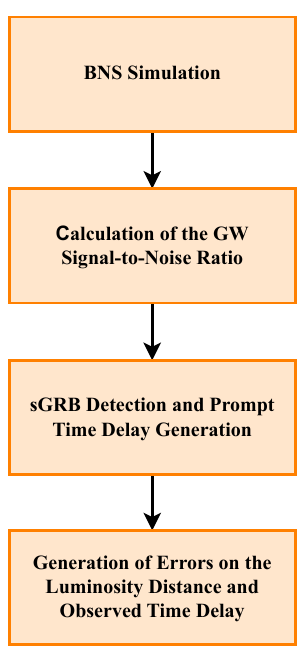}
    \caption{Flowchart depicting the simulation procedure to generate the GWs and sGRBs mock catalogs.}
    \label{fig:flowchart}
\end{figure}

In Fig.~\ref{fig:flowchart} we depict the simulation procedure used to generate the GW-sGRB mock catalogs.

\paragraph{Binary neutron star simulation:} We choose a flat \LCDM cosmology with parameters matching those derived from the latest observations of the cosmic microwave background by Planck \cite{2020} ($H_0 = 67.66 \, \hu$, $\Om = 0.310 $) and we also assume that GWs propagate at the speed of light ($\aO = 0$).
We simulate BNS sources with an isotropic distribution over the sky in terms of right ascension $\alpha$ and declination $\delta$. We also simulate an isotropic distribution for the binary orbital inclination angle $\iota$ and a uniform distribution of polarization angles $\psi$. 
The source masses $m_1$ and $m_2$ of the binary neutron stars are drawn uniformly in the interval $(1,3) \, \Msun$, with the condition $m_2 \leq m_1$. This distribution is consistent with the latest rate models inferred from the GW detections during the third LIGO-Virgo-KAGRA Observing Run (O3) \citep{theligoscientificcollaboration2022population}. The spin magnitudes of the two neutron stars are assumed to be vanishing. This is consistent with expectations for BNS, which are low-spin objects \cite{2019PhRvX...9a1001A}.
The distribution of the BNS sources in redshift is modeled according to a Madau-Dickinson-like star formation rate \citep{Madau_2014}, namely
\begin{equation}
\label{Madau_rate}
\frac{dN}{dV_c dt_s} \propto [ 1 + (1 + z_p)^{-\gamma - k}] \frac{(1 + z)^{\gamma}}{1 + \left(\frac{1+z}{1+z_{p}}\right)^{\gamma + k}}
\end{equation}
where the model parameters ($\gamma$, $k$, $z_p$) are set to (2.7, 6.0, 2.0). The parameters chosen for the BNS rate function are in agreement with the binary black hole merger rate inferred from O3 \cite{LIGOScientific:2020kqk,KAGRA:2021duu} as it is currently not possible to constrain the BNS rate as a function of redshift. GW events are simulated up to a maximum redshift that depends on the GW detector network configuration. We chose to simulate BNSs up to a maximum redshift 
$z_{\text{max}} = 0.2$ for O5, and $z_{\text{max}} = 2$ for ET. These values are high enough to embrace all sources detectable by the GW detectors with a joint sGRB observation.

\paragraph{Gavitational wave detection:} We consider a signal to be detected if its observed detector network signal-to-noise ratio (SNR) is greater than 12. This is a conservative threshold for the detection of GW events \cite{KAGRA:2013rdx}. The network SNR is defined as the quadrature sum of the individual detector SNRs 
\begin{equation}
\rho^2_{\text{net}}=\sum_{i \in {\rm IFO}} \rho^2_{\text{opt},i}\,.
\end{equation}
In this work, we model the GW SNR $\rho_{\text{opt},i}$ at the detector $i$ at the 0th post-Newtonian order \citep{Cutler_1994}, namely
\begin{equation} 
\label{eq:snr}
\rho^2_{\text{opt},i} = \frac{4A^2}{\Dl^2} \left[F^2_{+,i}( \alpha, \delta, \psi,t)(1 + \cos^2 \iota) + 4F^2_{\times,i}(\alpha, \delta, \psi,t)\cos^2 \iota\right] I_i\,,
\end{equation}
where the amplitude $A$ is given by
\begin{equation} \label{eq:amplitude}
A = \sqrt{\frac{5}{96}} \left(\frac{G\mathcal{M}_{\text{obs}}}{c^3}\right)^{5/6} c \pi^{-2/3}\,,
\end{equation}
with $\mathcal{M}_{\text{obs}}$ the redshifted chirp mass. $F_{+,i}$ and $F_{\times,i}$ are the detector antenna patterns for the detector $i$, for a GW signal located at right ascension $\alpha$ and declination $\delta$, with polarization angle $\psi$, and arrival at a time $t$. In our simulations, we fixed the antenna pattern functions at the arrival time $t$ of the GW. This approximation is reasonable for the current ground-based detector, as the BNS signals enter the sensitivity band around $10$ Hz, with only a few minutes left prior to the merger. Since the antenna patterns vary with the sidereal day, they can be considered constant for this case.
However, for ET, BNS signals will enter the sensitivity band around $\sim O(1)$ Hz and be visible for a few days. In order to ease the computational load of our simulations, we impose a lower limit to the ET sensitivity at $10$ Hz. In other words, we are artificially reducing the GW detection range that ET would have. However, as we will see later, this is not a problem for the simulation as the detection range for GRBs with Fermi-like satellites is significantly smaller than the restricted ET detection range. 

The quantity $I_i$ is defined as
\begin{equation} \label{eq:i7}
I_i = \int^{f_{\text{isco}}}_{10 \, \text{Hz}} \frac{f^{-7/3}}{S_{n,i}(f)} df\,,
\end{equation}
where $f_{\text{isco}}\approx 4.4 \, \text{kHz} \, (\Msun/M_{\text{tot}})$\footnote{$M_{\text{tot}}$ is the total mass of the BNS system at the observer.} is the orbital frequency of the innermost circular orbit of the BNS system, and $S_{n,i}(f)$ is the detector’s power spectral density. For this, we use the sensitivities expected to be reached by Advanced LIGO and Virgo during O5\footnote{\href{https://dcc.ligo.org/LIGO-T2000012-v1/public}{https://dcc.ligo.org/LIGO-T2000012-v1/public}} \cite{KAGRA:2013rdx} and by ET \cite{2023JCAP...07..068B}. Finally, we assume 100\% duty cycles for all detectors.

\paragraph{Short gamma-ray burst detection:} We approach the simulation of sGRB signals following \cite{Salafia_2015}. 
We assume that every BNS merger produces an sGRB prompted by a jet of relativistic matter ejected from the merger. We assume that the jet is narrowly collimated with an opening angle $\theta_B$ around the direction perpendicular to the binary orbital plane. We model the luminosity of the sGRB as
\begin{equation}
\label{eq:lum}
L(\iota; \theta_B, L_{\text{max}}) = L_{\text{max}}e^{-\frac{\iota^2}{2\theta_B^2}} 
\end{equation}
with $\theta_B=4.5\,$rad, compatible with the average value of the opening angle as found by \cite{Ghirlanda_2016}, and $L_{\text{max}}$ is drawn from a log-normal distribution with a mean of $5 \times 10^{51}$ erg s$^{-1}$ and width of 0.56 dex \citep{Salafia_2015}.
The simple parametrization in Eq.~\eqref{eq:lum} predicts a small luminosity if the binary is observed edge-on $\iota=\pi/2 \gg \theta_B$ and a luminosity equal to $L_{\text{max}}$ when the binary is observed face-on ($\iota=0\,$rad). This model is consistent with most of the observed sGRBs with known redshift and jet opening angle \cite{Fong_2013, Abbott_2017}.
To compute the sGRB peak energy in the source frame $E_{p}$, we model a relation between $E_{p}$ and $L(\iota; \theta_B, L_{\text{max}})$ in analogy to the one of long bursts \citep{Virgili2011}. Following \citep{Ghirlanda_2016}, we write this correlation in the form
\begin{equation}
\log_{10}\left(\frac{E_p}{670 \, \text{keV}}\right) = q + m \log_{10}\left(\frac{L}{10^{52} \, \frac{\text{erg}}{\rm s} }\right)\,,
\label{eq:Ep-Liso}
\end{equation}
the parameters $q$ and $m$ are set to 0.034 and 0.82, respectively.
For the detection of the sGRB, we assume a \textit{Fermi} gamma-ray burst monitor like experiment \citep{Thompson_2022}: if the sGRB peak flux surpasses the detection threshold of 10 $\frac{\text{ph}}{\text{s} \; \text{cm}^2}$ and $E_p$ falls within the 10–1000 keV energy band, we consider it to be detected. The spectra of sGRBs typically exhibit an observer frame peak energy distribution centered between 0.5 and 1 MeV. In this work, we assume that the entire luminosity is radiated at the energy of the peak flux $E_{p}$.

The next step is to draw a value of the GW-sGRB prompt time delay \Dts for each BNS merger to simulate the time of arrival of the GW and sGRB signals. Unfortunately, there is currently no robust model for this quantity, although the literature generally agrees that these should have values between $10^{-2}$ and $100$ seconds \citep{Zhang_2019}. Given these uncertainties, we use a simple Gaussian model to describe the distribution of prompt time time delays. Specifically, we consider three Gaussian distributions, each with a mean $\mu_{\Delta t}$ of 1.7 seconds and standard deviations $\sigma_{\Delta t}$ of 1.7, 0.17, and 0.017 seconds, respectively. We chose three different values of $\sigma_{\Delta t}$ as we expect that, if the distribution of prompt time delay were almost a constant in nature, it would be easier to obtain a statistical estimation of the sources' redshift. On the other hand the value of $\mu_{\Delta t}$ is set to be in agreement with the GW170817/GRB 170817A joint observations \cite{Abbott_2017}.

\paragraph{Error budgets on luminosity distance and observed time delays:} To represent measurement uncertainties inherent in actual observations, we generate simulated error values for the sources' luminosity distances and observed GW-sGRB time delays. 
The measurement of the luminosity distance is provided from the GW detection. In principle, to obtain reliable distance estimates, one should use a full Bayesian analysis using simulated data and implementing state-of-the-art waveform models for the GW signals. However, this approach is highly computationally demanding. For this reason, we adopt a GW likelihood \citep{Gair_2023} model able to describe statistically the typical error budgets for the GW luminosity distance. With this model, we associate a set of data $x_i$ with an ``observed luminosity distance'' $D_L^{\text{obs}}$, namely
\begin{equation} \label{eq:lum_dist_likelihood}
\mathcal{L}_{\text{noise}}(D_L^{\text{obs}}|D_L^{\text{true}}) = \mathcal{N}(D_L^{\text{obs}}|\mu = D_L^{\text{true}}, \sigma = 0.2 \cdot D_L^{\text{true}})\,,
\end{equation}
where $D_L^{\text{true}}$ is the true physical value of the luminosity distance. 
Additionally, we also assume that the observed SNR is not the true SNR in Eq.~\eqref{eq:snr} but it is modified by possible noise fluctuations. To mimic this effect, we generate the observed SNR $\rho_{\text{obs}}$ from a Gaussian distribution with a mean equal to the intrinsic SNR $\rho_{\text{true}}$ and a standard deviation of 1 \cite{Fishbach:2018edt} and apply the SNR selection criterion on it. The likelihood for the observed SNR is 
\begin{equation} \label{eq:snr_likelihood}
\mathcal{L}_{\text{noise}}(\rho_{\text{obs}}|\rho_{\text{true}}) = \mathcal{N}(\rho_{\text{obs}}|\mu = \rho_{\text{true}}, \sigma = 1)\,.
\end{equation}
For the observed time delays $\Delta t_d$, we sample the noisy measurement from a Gaussian distribution, 
\begin{equation} \label{eq:time_delay_likelihood}
\mathcal{L}_{\text{noise}}(\Delta t_d^{\text{obs}}|\Delta t_d^{\text{true}}) = \mathcal{N}(\Delta t_d^{\text{obs}}|\mu = \Delta t_d^{\text{true}}, \sigma = 0.05\,\text{s})\,,
\end{equation}
where the standard deviation is calibrated on the precision achieved for the timing of GW170817 and GRB 170817A \cite{2017ApJ...848L..12A}. 
The overall likelihood model is given by the product of Eqs.~\eqref{eq:lum_dist_likelihood}--\eqref{eq:time_delay_likelihood}, the posterior estimation values of the luminosity distance and time delays are drawn sampling from this likelihood.

\subsection{Catalogs description}
\label{sec:3b}

\begin{figure}
    \centering
    \includegraphics[scale=0.4]{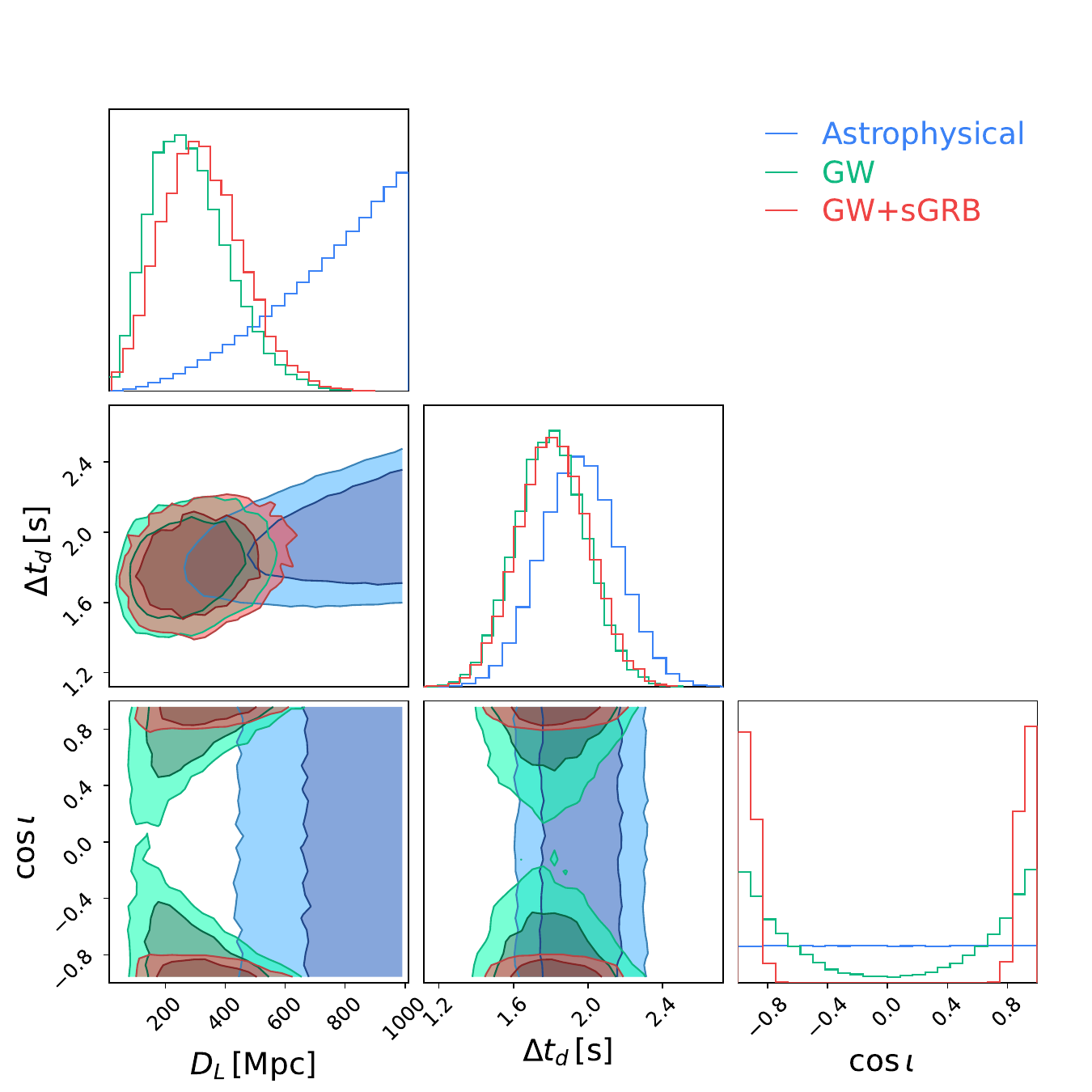}
    \caption{O5 catalog: 
    normalized distribution of the luminosity distance \Dl, cosine of the inclination angle $\cos\iota$, and observed time delay \Dtd for Astrophysical events (blue line), GW detections (green line), and joint GW-sGRB detections (red line). The plots report the $68.3\%$ and $90\%$ credible intervals.}
    \label{fig-2}
\end{figure}

\begin{figure}
    \centering
    \includegraphics[scale=0.4]{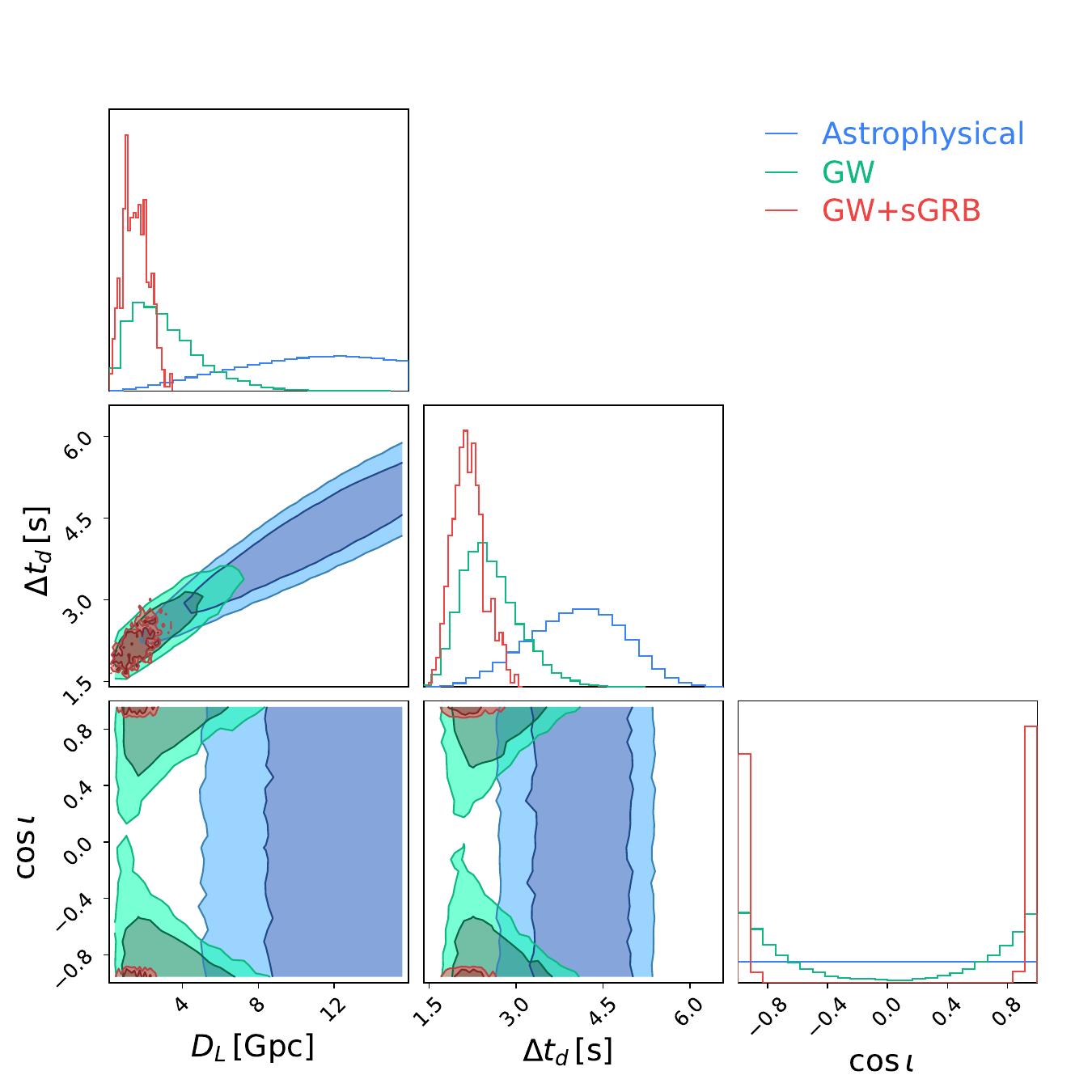}
    \caption{ET catalog: normalized distribution of the luminosity distance \Dl, cosine of the inclination angle $\cos\iota$, and observed time delay \Dtd for Astrophysical events (blue line), GW detections (green line), and joint GW-sGRB detections (red line). The plots report the $68.3\%$ and $90\%$ credible intervals.}
    \label{fig-3}
\end{figure}

We describe the population distribution of the mock catalogs, focusing on the luminosity distance \Dl, the inclination angle $\iota$, and the detected time delay \Dtd. Figures \ref{fig-2} and \ref{fig-3} show the distributions obtained for \Dl, \Dtd an $\cos \iota$ for the cases of O5 and ET. For these example cases, we simulate $10^6$ BNS mergers with a standard deviation of the Gaussian prompt time delay distribution set to $\sigma_{\Delta t} = 0.17\,$s.


 The distributions of detected BNS mergers feature two peaks at $\cos\iota \sim -1$ and $\cos\iota\sim 1$. This is the result of the dependence of the GW SNR and the sGRB intrinsic luminosity on the inclination angle $\iota$ [see Eqs.~\eqref{eq:snr} and \eqref{eq:lum}], which causes the detection probability to be highest when the BNS merger orbital angular momentum is aligned or antialigned with respect to our line of sight (face-on/face-away). While GW detections span all $\cos\iota$ values, no sGRBs are detected within the range $\cos\iota \in (-0.70,0.70)$. This is a consequence of the exponential dependency of the sGRB luminosity on the inclination angle [see Eq.~\eqref{eq:lum}].
 
We detect GWs up to luminosity distances $D_L \sim 900 $ Mpc for O5 and $D_L \sim 15000 \; \text{Mpc}$ for ET. For the O5 case, sGRBs will be detectable for almost all GW signals nearly face-on ($|\cos\iota|>0.70$). However, for ET, the GW detection reach will greatly surpass the sGRB detection reach and we will be able to detect sGRBs only for face-on binaries below $D_L \sim 3500 \; \text{Mpc}$. Consequently, about $\sim44\%$ of BNS detections in O5 are followed by an sGRB observation, and this rate drops to just $\sim4\%$ for ET (even though in terms of absolute numbers GW-sGRB joint detections in ET will be more abundant) \cite{2022A&A...665A..97R}.
An increase in luminosity distance detection range correlates with a shift and a broadening in the observed time delay $\Delta t_{d}$ distribution as may be clearly seen by comparing Figs.~\ref{fig-2} and \ref{fig-3}. This is a consequence of their shared dependency on the merger redshift, $z$ [refer to Eqs.~\eqref{eq:Dl} and \eqref{eq:time_delay}]. The $D_L$ - $\Delta t_d$ distribution could then be used to extract implicit redshift information from a set of detected time delays and luminosity distances. However, for BNS merger events detected in O5, the observed time delay distribution does not evolve significantly with luminosity distance. In such scenario, the redshift effect is very small and, as we will show later, it will be difficult to measure it.

\section{Results}
\label{sec:4}

Using the catalogs of GW and sGRB detections, we forecast the precision with which the prompt time delay distribution, speed of gravity and $H_0$ can be constrained thanks to time-delay cosmography. To accomplish this, we use \icarogw a Python tool developed to infer the population properties of CBCs observed through GWs \citep{mastrogiovanni2023icarogw}. More precisely, we employ a Markov Chain Monte Carlo algorithm \footnote{The \textsc{emcee} sampling algorithm \cite{2013PASP..125..306F} from \textsc{Bilby} \cite{bilby_paper,bilby_pipe_paper}}, to sample the population parameters.
We consider three inference scenarios:
\begin{enumerate}[label=\textnormal{\Roman*.}]
    \item \textit{Speed of gravity and prompt time delays}. We fix $H_0$ to the Planck value, i.e., $H_0 = 67.66 \, \hu$ \citep{2020}, and jointly infer the speed of gravity and prompt time delay distribution with a set of 3 parameters ($\hat{\alpha}_0,\mu_{\Delta t},\sigma_{\Delta t}$).
    \item \textit{Cosmology and prompt time delays}. We set $\hat{\alpha}_0=0$ (GR value), and jointly infer the cosmology ($H_0$) and prompt time delay distributions ($\mu_{\Delta t},\sigma_{\Delta t}$). 
    \item \textit{Fully agnostic}. We jointly infer all four parameters ($\hat{\alpha}_0,H_0,\mu_{\Delta t},\sigma_{\Delta t}$).
\end{enumerate}
In each scenario, we consider sets of $N_{\text{obs}} = [4,10,100]$ GW-sGRB detections. The choice to start with $N_{\text{obs}} = 4$ detections is motivated by the fact that in scenario III, where we jointly infer 4 parameters, it would be impossible to jointly constrain all parameters with less than 4 observations. In each scenario, we also consider multiple catalogs constructed for the O5 and ET sensitivities, and with various standard deviations of the prompt time delay $\Delta t_s$ distributions, namely $\sigma_{\Delta t} = [1.7,0.17,0.017]$ s.

The priors that we set for the population parameters are summarized in Table~\ref{table:my_label}. For the prompt time delay distribution, we use a uniform prior between $[0, 5]$ s for the mean $\mu_{\Delta t}$ and a flat-in-log prior between $[10^{-3},10]$ s for the standard deviation $\sigma_{\Delta t}$. The choice to use a flat-in-log prior within this range stems from the different $\sigma_{\Delta t}$ values used for the $\Delta t_s$ distribution models: the flat-in-log prior guarantees a uniform distribution across all these orders of magnitude. For the Hubble constant, we choose a uniform prior distribution between $[30, 140] \, \hu$. Finally, for $|\hat{\alpha}_0|$ we use a uniform prior with a maximum value dependent on the detection and simulation scenario, spanning between $\sim{10^{-19}}$ and $\sim{10^{-15}}$. The different prior range is chosen so that the posterior on \sg is fully contained in the prior range. The values of the remaining population parameters are fixed in all scenarios. In the remainder of this work, any references to ``uncertainty'' and ``precision'' correspond to the median with symmetric 68.3\% credible interval.
\begin{table}[!htb]
\centering
\caption{List of priors for the population parameters.}
\begin{tabular}{@{}cccc@{}}
\toprule[1.pt]
\toprule[1.pt]
\addlinespace[0.3em]
\textbf{Parameter} & \textbf{True Value} & \thead{\textbf{Prior} \\ \textbf{distribution}} & \thead{\textbf{Distribution} \\ \textbf{range}} \\
\midrule
$\mu_{\Delta t}$ & $1.7 s$ & Uniform & [0,5] s \\
\addlinespace[0.2em]
$\sigma_{\Delta t}$ & [1.7,0.17,0.017] s & logUniform & [$10^{-3}$,10] s\\
\addlinespace[0.2em]
$H_0$ & $67.66 \, \hu$ & Uniform & $[30,140] \, \hu$ \\
\addlinespace[0.2em]
$\hat{\alpha}_0$ & 0 & Uniform & Variable \\
\addlinespace[0.2em]
$\Omega_M$ & 0.310 & Fixed &  \\
\addlinespace[0.2em]
$\gamma$ & 2.7  & Fixed & \\
\addlinespace[0.2em]
k & 6 & Fixed & \\
\addlinespace[0.2em]
$z_p$ & 2 & Fixed & \\
\bottomrule[1.pt]
\bottomrule[1.pt]
\end{tabular}
\label{table:my_label}
\end{table}

\subsection{Scenario I: Speed of gravity and prompt time delays}
\label{SCEN1}
Scenario I corresponds to the case in which we are interested in constraining only the speed of gravity and the prompt time delay distribution. We find that future observations will allow us to jointly constrain the speed of gravity and the prompt time delay distribution.

\begin{figure}[!tb]
    \centering
    \includegraphics[width=0.5\textwidth]{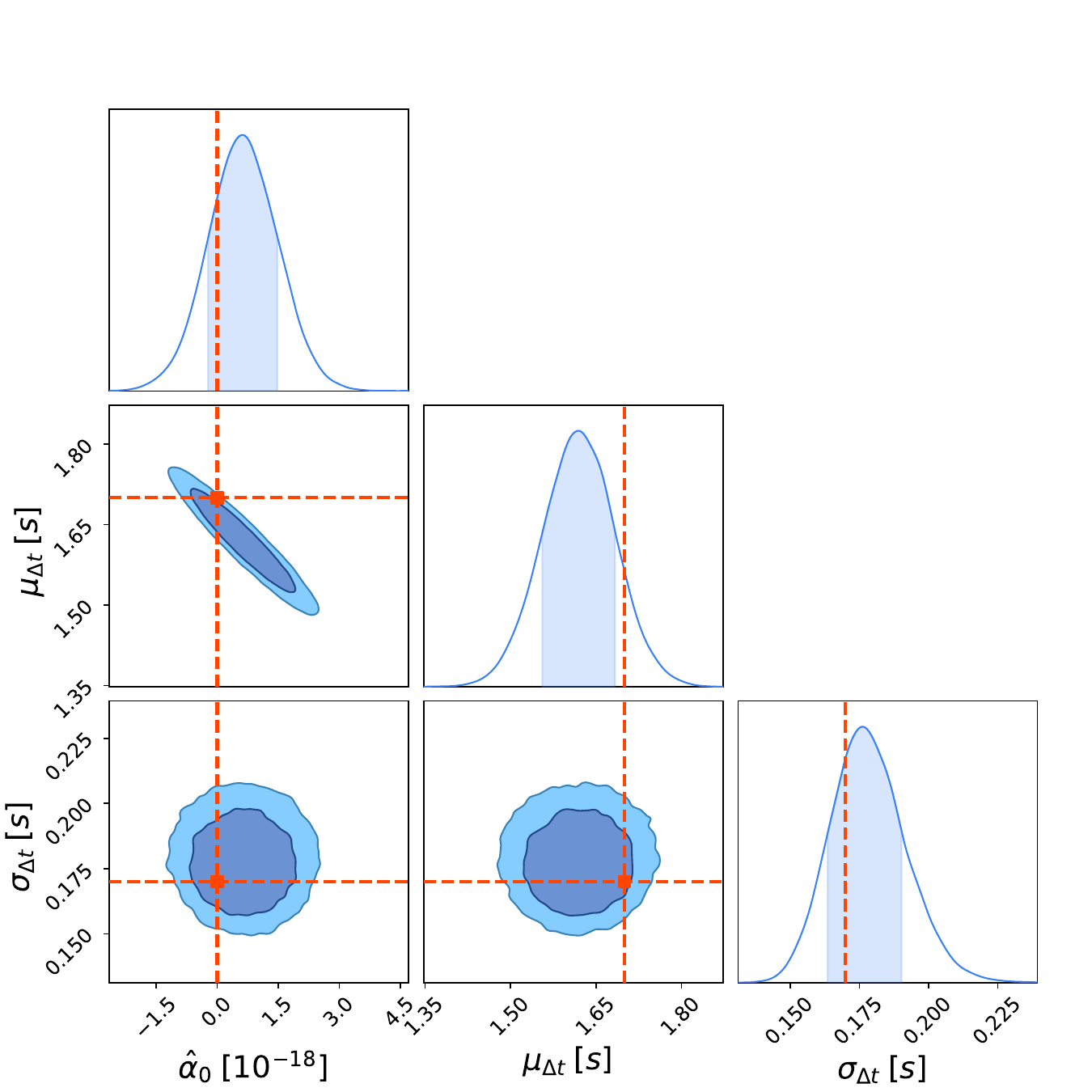}
    \caption{Joint and marginalized posterior distributions for $\hat{\alpha}_0$, $\mu_{\Delta t}$ and $\sigma_{\Delta t}$ in the case of 100 events detected by ET and the true value of $\sigma_{\Delta t} = 0.17$ s. The orange lines represent the true value of the parameters. The plots report the $68.3\%$ and $90\%$ credible intervals. The marginal 1-dimensional distribution reports the $68.3\%$ credible intervals.}
    \label{fig.20}
\end{figure}
\begin{figure}[!htb]
    \centering
    \includegraphics[width=0.425\textwidth]{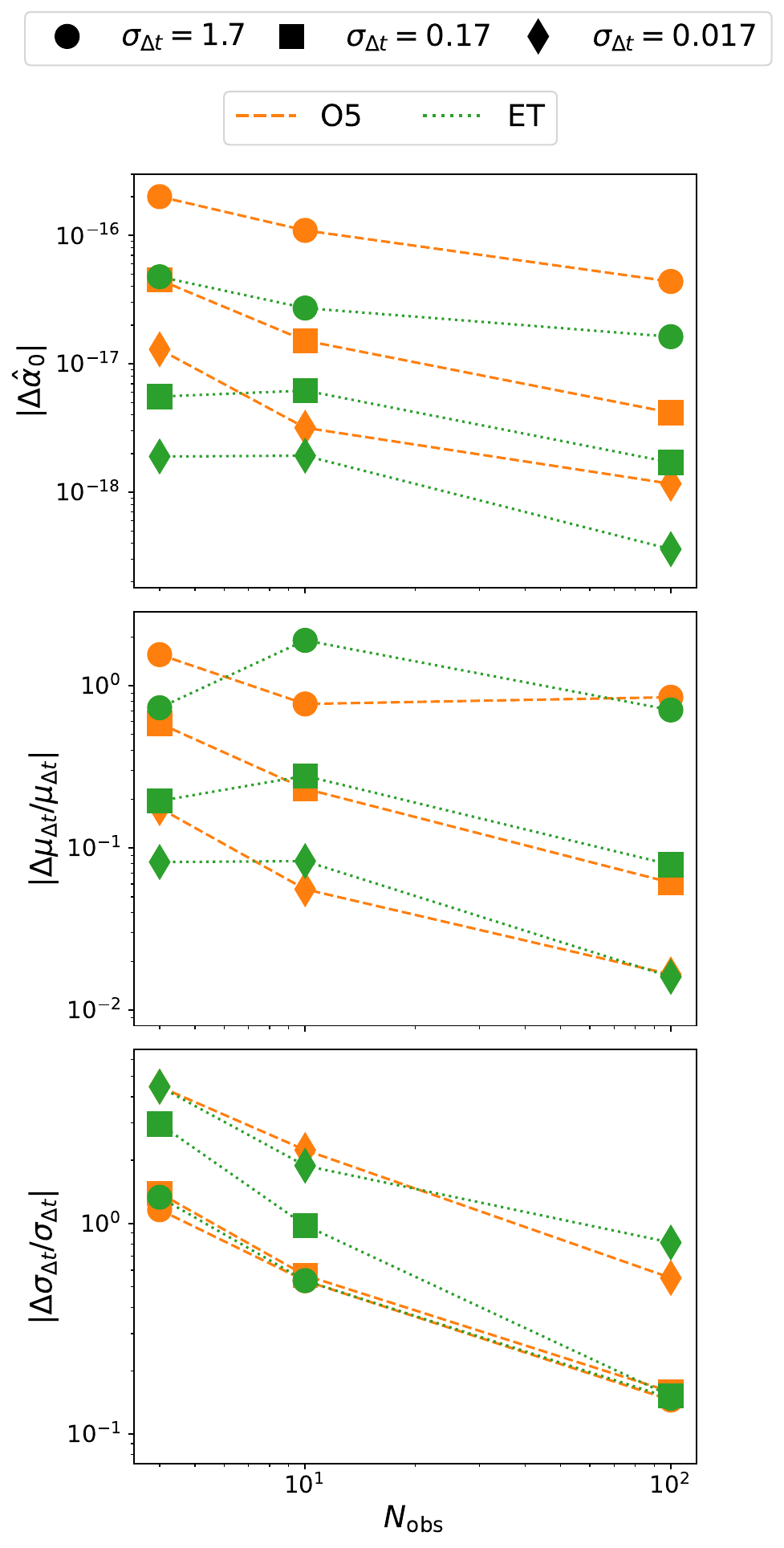}
    \caption{Precision at the 68.3\% credible interval for scenario I on the population parameters when we combine more and more joint GW-sGRB detections. Each row of subplots represents a deduced population parameter $(\hat{\alpha}_0,\mu_{\Delta t},\sigma_{\Delta t})$. Each subplot shows 6 curves; one for each prompt time delay model ($\sigma_{\Delta t} = [1.7,0.17,0.017]$ s) and configuration of the detector network (O5, ET).}
    \label{fig.37}
\end{figure}

As an example, in Fig.~\ref{fig.20}, we present the joint posterior distribution for 100 events detected by ET. 
The interplay between source-frame time delay parameters and the speed of gravity is evident in the strong correlation between $\hat{\alpha}_0$ and $\mu_{\Delta t}$ exhibited by the posterior distribution. Considering the expression of the detected time delay in Eq.~\eqref{eq:time_delay}, a lower $\hat{\alpha}_0$ must be compensated by a higher source-frame time delay to maintain the observed time delay value fixed at its detected value. 
Figure \ref{fig.20} clearly shows that the GW-sGRB prompt time delay distribution and speed of gravity can be measured jointly.
In Fig.~\ref{fig.37} we summarize the precisions achievable for the population parameters $(\hat{\alpha}_0,\mu_{\Delta t},\sigma_{\Delta t})$. 
In our simulations, we successfully constrained the speed of gravity across all observation counts and prompt time delay models with an uncertainty $\Delta \hat{\alpha}_0$ ranging from $\sim 2\cdot 10^{-16}$ down to $\sim 10^{-19}$. These constraints are up to $10^4$ times better than those derived from the GW170817 measurement \cite{Abbott_2017}. We also notice that the precision on $\hat{\alpha}_0$ depends on the GW detection range and prompt time delay model ($\sigma_{\Delta t} = [1.7, 0.17,0.017]\,$s) as well as the number of detected GW events $N_{\text{obs}}$. In Fig.\ref{fig.37}, for a fixed detector sensitivity and prompt time delay distribution model, the precision in inferring the parameters can decrease from 4 to 10 observations. This behavior arises due to statistical fluctuations, as different event realizations lead to variations in the precision of the inferred parameters. The scatter is more significant with fewer observations. In fact, with a sufficiently large number of events, the uncertainty is expected to scale as $1/\sqrt{\Nobs}$, reducing the impact of these fluctuations.
We find that a ``sharper" prompt time delay $\Delta t_s$ distribution, i.e., $\sigma_{\Delta t}$ is 10 (100) times smaller than $\sigma_{\Delta t} = 1.7\,$s, decreases the uncertainty of an $\hat{\alpha}_0$ measurement by a factor $\sim$ 0.1 (0.035).
Meanwhile, we find that using more sensitive detectors will improve the constraints on $\hat{\alpha}_0$. Specifically, ET will be able to constrain $\hat{\alpha}_0$ 3 times better than O5-like sensitivities.
The measurements of the prompt time delay mean $\mu_{\Delta t}$ and standard deviation $\sigma_{\Delta t}$ are constrained with an uncertainty of $\sim 0.6-0.015\,$s and $\sim 0.5-0.006\,$s respectively. Similar to $\hat{\alpha}_0$, the precision of these constraints depends on both $N_\text{obs}$ and the prompt time delay model scenario. However, $\mu_{\Delta t}$ could only be constrained with a minimum of 100, 10 and 4 observations for a prompt time delay distribution with $\sigma = 1.7$\,s, 0.17\,s, and 0.017\,s, respectively. To constrain $\sigma_{\Delta t}$ at least 10 observations were necessary for $\sigma_{\Delta t} = 1.7\,$s, while 4 observations sufficed for both 0.17\,s and 0.017\,s.
The uncertainty of the ($\mu_{\Delta t}$, $\sigma_{\Delta t}$) measurements varies with the “sharpness” of the prompt time delay distribution in the same qualitative way as in the $\hat{\alpha}_0$ case.
From Fig.~\ref{fig.37}, we conclude that the constraints on the prompt time delay population parameters are less dependent on the GW detector sensitivities than the $\hat{\alpha}_0$ constraints. This is consistent with the expected error budgets from Eq.~\eqref{eq:time_delay}. Assuming Gaussian errors for $\Delta t_d$, and a known prompt time delay distribution, we can see that the expected standard deviation on \aO is inversely proportional to the lookback time \Tl. Repeating the same procedure for the prompt time delay but fixing $\aO=0$, provides a precision on \Dts inversely proportional to $1+z$. As the lookback time \Tl grows faster than $1+z$, for high redshift values (higher detection ranges), the precision on \aO will improve faster than the ones on prompt time delay distributions.

\subsection{Scenario II: Cosmology and prompt time delays}
\label{SCEN2}

Scenario II is the case in which we would like to exploit time delays to measure $H_0$ when the source redshift is unknown. 

\begin{figure}[!htb]
    \centering
    \includegraphics[width=0.425\textwidth]{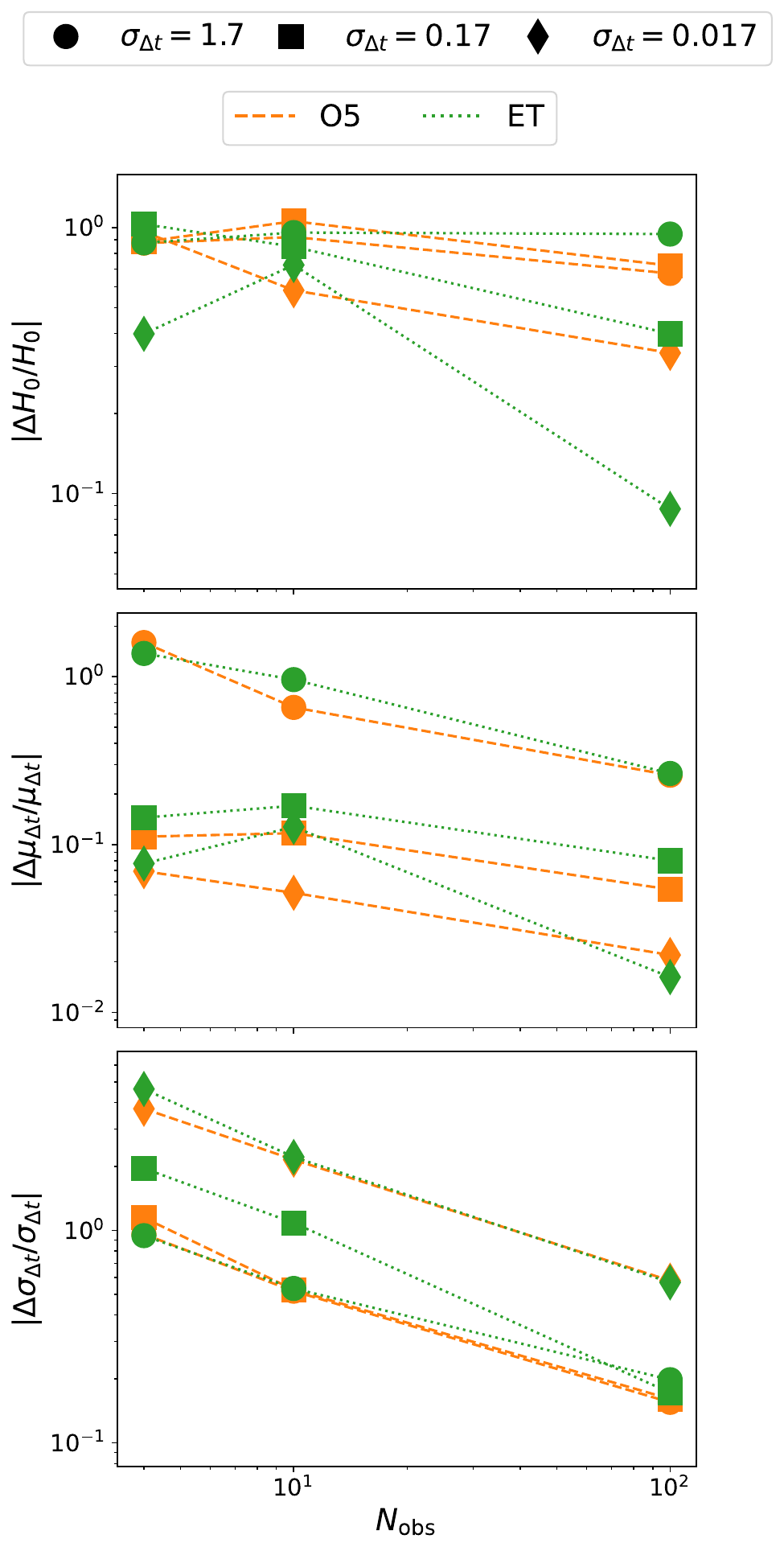}
    \caption{Precision at the 68.3\% credible interval for scenario II on the population parameters when we combine more and more joint GW-sGRB detections. Each row of subplots represents a deduced population parameter $(H_0,\mu_{\Delta t},\sigma_{\Delta t})$. Each subplot shows 6 curves; one for each prompt time delay model ($\sigma_{\Delta t} = [1.7,0.17,0.017]$ s) and configuration of the detector network (O5, ET).}
    \label{fig.38}
\end{figure}

Figure \ref{fig.38} shows the precision on $H_0$ and the prompt time delay parameters that we could be able to achieve with this technique.
In general, we observe that the constraining power on $H_0$ increases as the detector network sensitivity improves and the prompt time delay distribution becomes ``sharper'' (lower $\sigma_{\Delta t}/\mu_{\Delta t}$). For a GW detector network with O5-like sensitivities, $H_0$ can be constrained with a precision $<30\%$ when 100 sources will be detected and the prompt time delays have a narrow distribution ($\sigma_{\Delta t}=0.017\,$s). If the distribution of prompt time delays is wider, with an O5-like detector network, we will not be able to constrain $H_0$ even with 100 of these sources. Instead, for ET, which will observe GW events at higher redshifts, we find that $H_0$ could be measured already at $\sim 30\%$ precision with 100 sources and a distribution of prompt time delays that is mildly spread ($\sigma_{\Delta t}=0.17\,$s). In the ET case, we obtain that with 100 of these sources, it will be possible to constrain $H_0$ at $<10\%$ precision, which would be enough to provide useful hints on the $H_0$ tension. In the best case scenario, for $N_{\text{obs}} = 100$, ET and $\sigma_{\Delta t} = 0.017$ s, we obtain a value for the Hubble constant of $H_0 = 63.69^{+2.81}_{-2.71} \hu$. 
For the prompt time delay parameters, by comparing Figs.~\ref{fig.38} and Fig.~\ref{fig.37}, we find that the qualitative behavior for these two parameters is similar to what was found in scenario I.

\begin{figure}[!tb]
    \centering
    \includegraphics[width=0.5\textwidth]{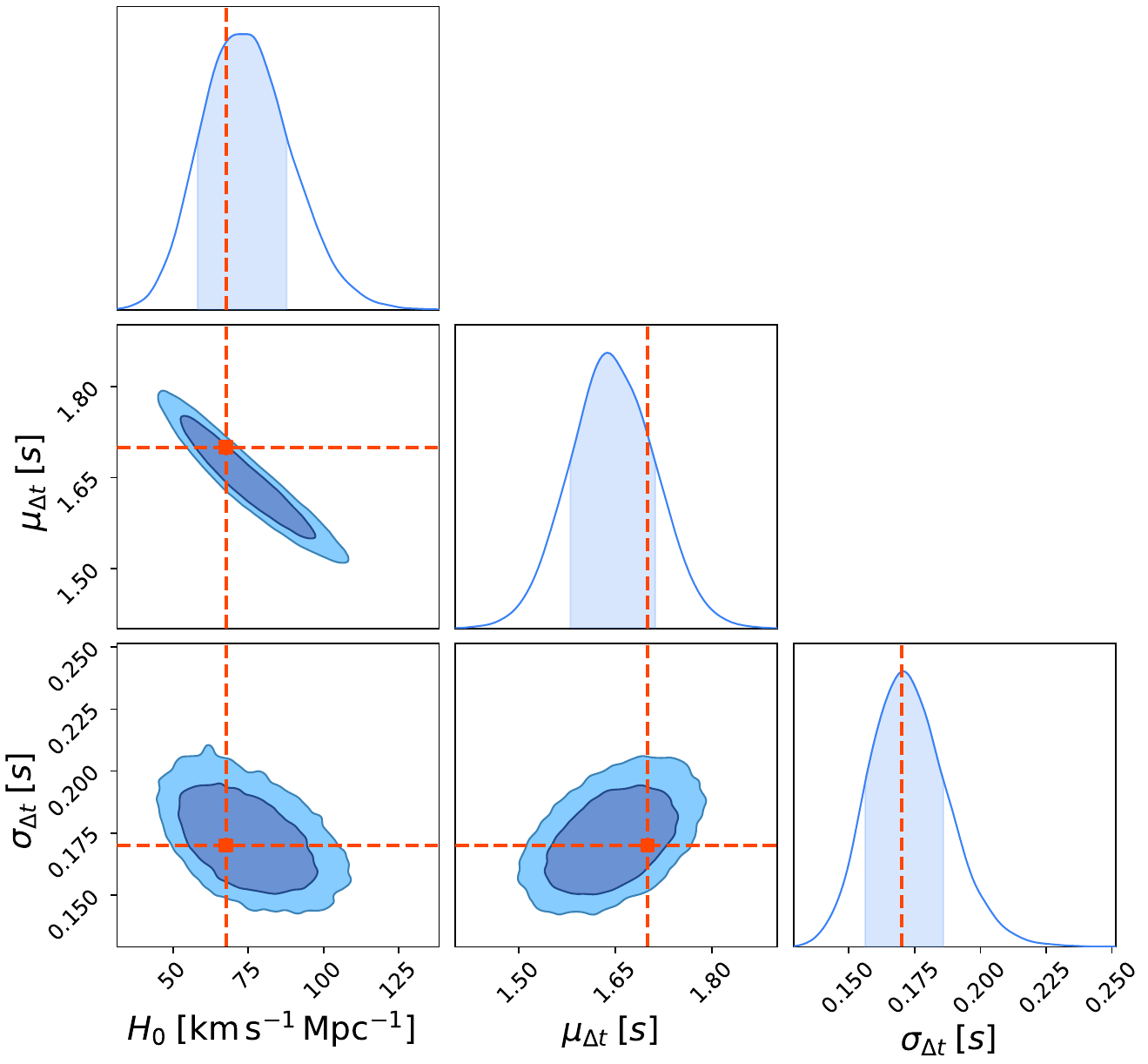}
    \caption{ Joint and marginalized posterior distributions for $H_0$, $\mu_{\Delta t}$ and $\sigma_{\Delta t}$ in the case of 100 events detected by ET and true value of $\sigma_{\Delta t} = 0.17$ s. The orange lines represent the true value of the parameters. The plots report the $68.3\%$ and $90\%$ credible intervals. The marginal 1-dimensional distribution reports the $68.3\%$ credible intervals.}
    \label{fig.21}
\end{figure}

In Fig.~\ref{fig.21}, we show the joint posterior distribution for 100 events detected by ET and a mildly narrow prompt time delay distribution ($\sigma_{\Delta t}=0.17\,$s). 
The posterior distribution shows a strong correlation between the measurement of $H_0$ and $\mu_{\Delta t}$. As can be seen from Eq.~\eqref{eq:time_delay}, since lower $H_0$ values lead to lower redshifts, $\mu_{\Delta t}$ must increase to keep $\Delta t_{d}$ fixed to its detected value.
We also observe a weak correlation for the pairs ($H_0$, $\sigma_{\Delta t}$) and ($\mu_{\Delta t}$, $\sigma_{\Delta t}$). In fact, a larger $H_0$ results in a larger spread of the BNS redshifts. Then, due to Eq.~\eqref{eq:time_delay}, an increase in $H_0$ also widens the spread of $\Delta t_{d}$ beyond the range actually observed. To reconcile this, the standard deviation of the prompt time delays $\sigma_{\Delta t}$ needs to be reduced. Then $H_0$ is correlated with both $\mu_{\Delta t}$ and $\sigma_{\Delta t}$: this induces a correlation between them in the marginalized posterior distribution.

\subsection{Scenario III: Fully agnostic}
\label{SCEN3}

In this scenario, we take an agnostic approach and assume that we wish to measure prompt time delay distributions, $H_0$ and also the speed of gravity. 

\begin{figure}[!tb]
    \centering
    \includegraphics[width=0.425\textwidth]{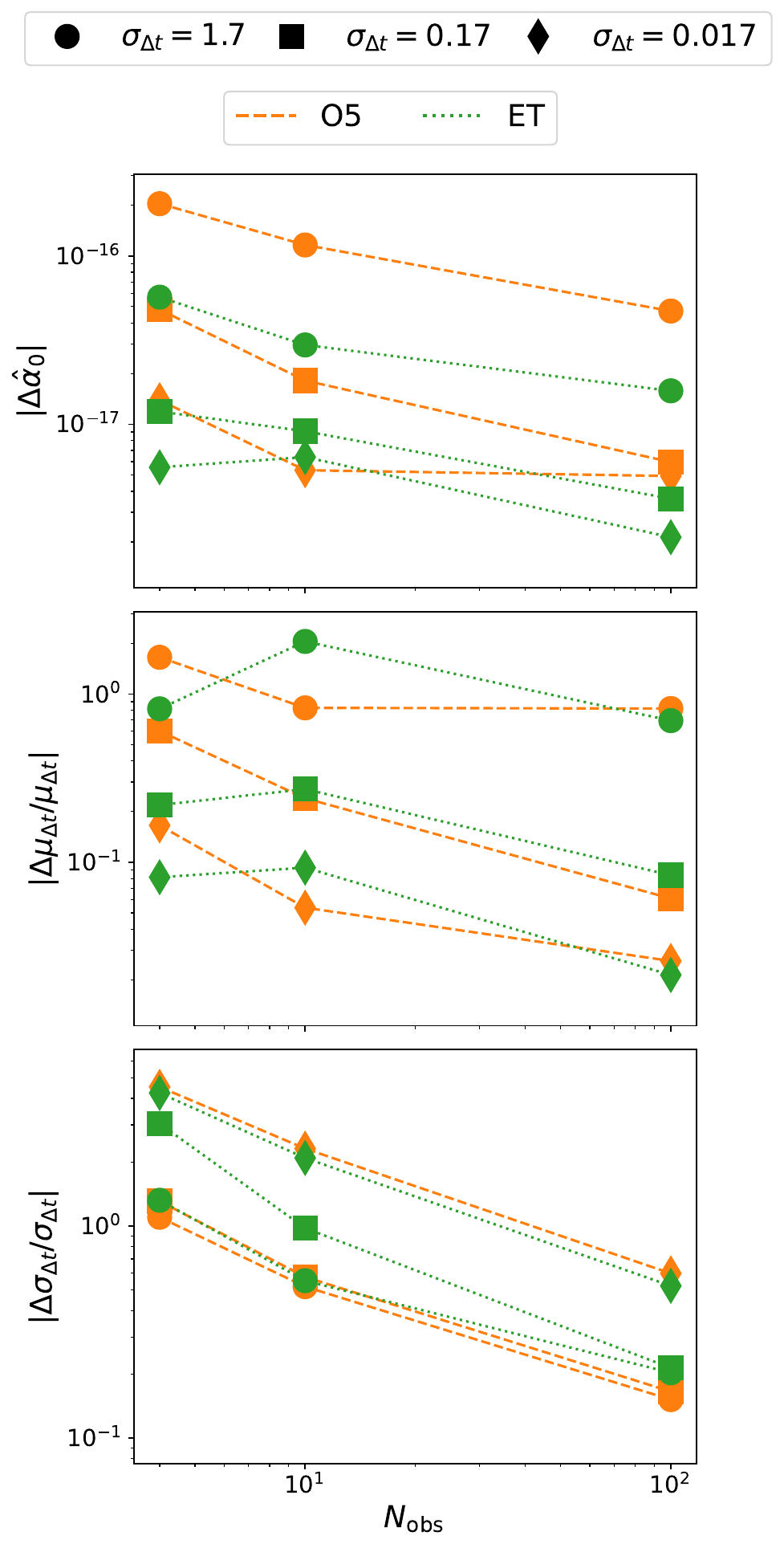}
    \caption{Precision at the 68.3\% credible interval for scenario III on the population parameters when we combine more and more joint GW-sGRB detections. Each row of subplots represents a deduced population parameter $(\hat{\alpha}_0,\mu_{\Delta t},\sigma_{\Delta t})$. Each subplot shows 6 curves; one for each prompt time delay model ($\sigma_{\Delta t} = [1.7,0.17,0.017]$ s) and configuration of the detector network (O5, ET). We do not display \HO as we are unable to constrain it for such inference scenario.}
    \label{fig.39}
\end{figure}

In this case, we find that it will be difficult to constrain the $H_0$ inside the prior range used for the analysis even when considering narrow prompt time delay distributions and a hundred sources. Figure \ref{fig.39} displays the uncertainties on the speed of gravity and prompt time delay parameters that we can constrain in this case.
The parameters $\hat{\alpha}_0$, $\mu_{\Delta t}$ and $\sigma_{\Delta t}$ are constrained with uncertainty of $\sim 2\cdot 10^{-16}-10^{-18}$, $\sim 0.5-0.02\,$s and $\sim0.5-0.006\,$s, respectively. In comparison to scenario I (cf.~Fig.~\ref{fig.37}), these parameters are less constrained. This is a consequence of the fact that in this case we are also fitting for $H_0$, which introduces additional degeneracies in the estimation of prompt time delay distributions.

\begin{figure}[!tb]
    \centering
    \includegraphics[width=0.5\textwidth]{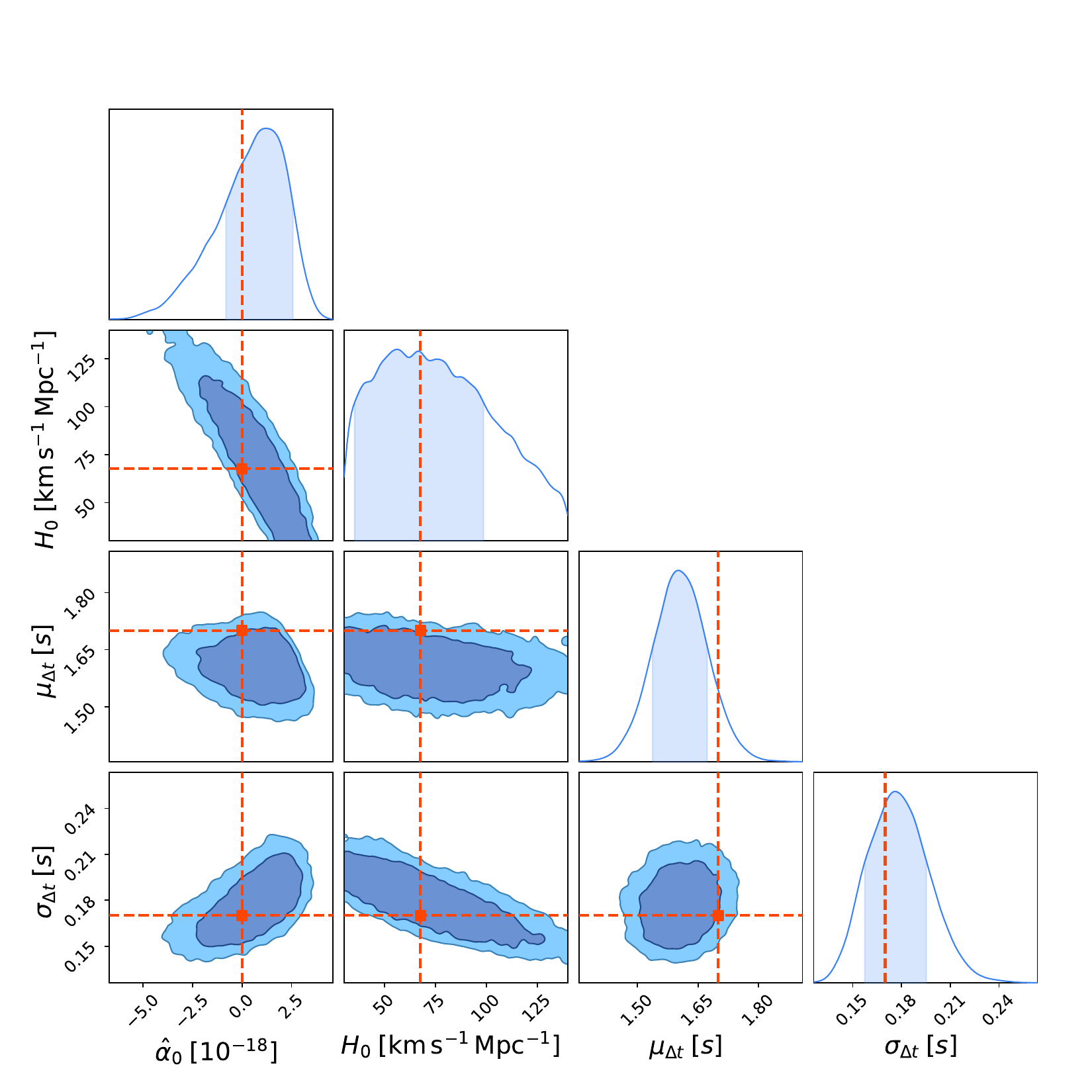}
    \caption{Joint and marginalized posterior distributions for $\hat{\alpha}_0$, $H_0$, $\mu_{\Delta t}$ and $\sigma_{\Delta t}$ in the case of 100 events detected by ET and the true value of $\sigma_{\Delta t} = 0.17$ s. The orange lines represent the true value of the parameters. The plots report the $68.3\%$ and $90\%$ credible intervals. The marginal 1-dimensional distribution reports the $68.3\%$ credible intervals.}
    \label{fig.22}
\end{figure}

For $H_0$, we find that, for each detector and prompt time delay model, we will not be able to measure the Hubble constant even with 100 GW detections. This is a consequence of the fact that the determination of $\hat{\alpha}_0$ is degenerate with the prompt time delay distribution, which is a crucial ingredient to get the implicit redshift information required for $H_0$. In other words, $H_0$ and the GW propagation speed are correlated with each other. The two parameters are degenerate since different combinations of their values can result in the same observed time delay. We display the correlations among the population parameters in Fig.~\ref{fig.22}, where we show the joint posterior distribution for 100 events detected by ET.

\section{Conclusion}
\label{sec:5}
In this paper, we presented a new methodology to jointly estimate the speed of gravity, the universe expansion, and prompt time delay distribution using joint GW-sGRB observations without the need for a direct redshift measurement.

In Sec.~\ref{sec:3}, we described the procedure followed to generate mock catalogs of joint GW-sGRB detections. We performed extensive simulations, considering two different detection networks (O5 and ET), three prompt time delay models $\Delta t_s$ ($\sigma_{\Delta t} = [1.7,0.17,0.017]\,$s), and three different observing scenarios with 4, 10, and 100 joint GW-sGRB detections. In Sec.~\ref{sec:4}, we used these catalogs to jointly infer the speed of gravity, the prompt time delay distribution, and the Hubble constant.

We found that constraining the Hubble constant through our method requires $\sim O(100)$ GWs-sGRB detections that will probably be observed with next-generation detectors. For instance, with 100 events detected by ET and standard deviation $\sigma_{\Delta t} = 0.17\,$s, we constrained the 68\% credible interval of $H_0$ to $ 74^{+15}_{-14} \hu$. We think this is not a viable method to measure the Hubble constant for O5 as we will have too few GWs-sGRBs events.

Constraints on the speed of gravity and prompt time delay distribution can be measured even with $\sim 10$ events, however. With 10 detections observed during O5 and a true value of $\sigma_{\Delta t} = 0.17\,$s, we set a constraint on $\hat{\alpha}_0$ of order $O(10^{-17})$ (100 times better than the current measurement from GW170817) while also constraining the mean and standard deviation of the $\Delta t_s$ distribution to $ 2.08^{+0.23}_{-0.24}$ s and $ 0.21^{+0.07}_{-0.05}$ s precision, respectively.

Finally, in the fully agnostic scenario (joint inference of all four parameters), we were unable to constrain $H_0$ for any of the GW-sGRB catalogs considered in this work.

\begin{acknowledgments}

The authors acknowledge the support of the computing facilities at INFN Rome of the Amaldi Research center funded by the MIUR program
“Dipartimento di Eccellenza” (CUP: B81I18001170001).
This material is based upon work supported by NSF's LIGO Laboratory which is a major facility fully funded by the National Science Foundation.

\end{acknowledgments}

\appendix

\newpage
\clearpage

\onecolumngrid

\newpage
\clearpage
\twocolumngrid


\bibliography{apssamp}

\begin{thebibliography}{57}%
\makeatletter
\providecommand \@ifxundefined [1]{%
 \@ifx{#1\undefined}
}%
\providecommand \@ifnum [1]{%
 \ifnum #1\expandafter \@firstoftwo
 \else \expandafter \@secondoftwo
 \fi
}%
\providecommand \@ifx [1]{%
 \ifx #1\expandafter \@firstoftwo
 \else \expandafter \@secondoftwo
 \fi
}%
\providecommand \natexlab [1]{#1}%
\providecommand \enquote  [1]{``#1''}%
\providecommand \bibnamefont  [1]{#1}%
\providecommand \bibfnamefont [1]{#1}%
\providecommand \citenamefont [1]{#1}%
\providecommand \href@noop [0]{\@secondoftwo}%
\providecommand \href [0]{\begingroup \@sanitize@url \@href}%
\providecommand \@href[1]{\@@startlink{#1}\@@href}%
\providecommand \@@href[1]{\endgroup#1\@@endlink}%
\providecommand \@sanitize@url [0]{\catcode `\\12\catcode `\$12\catcode `\&12\catcode `\#12\catcode `\^12\catcode `\_12\catcode `\%12\relax}%
\providecommand \@@startlink[1]{}%
\providecommand \@@endlink[0]{}%
\providecommand \url  [0]{\begingroup\@sanitize@url \@url }%
\providecommand \@url [1]{\endgroup\@href {#1}{\urlprefix }}%
\providecommand \urlprefix  [0]{URL }%
\providecommand \Eprint [0]{\href }%
\providecommand \doibase [0]{https://doi.org/}%
\providecommand \selectlanguage [0]{\@gobble}%
\providecommand \bibinfo  [0]{\@secondoftwo}%
\providecommand \bibfield  [0]{\@secondoftwo}%
\providecommand \translation [1]{[#1]}%
\providecommand \BibitemOpen [0]{}%
\providecommand \bibitemStop [0]{}%
\providecommand \bibitemNoStop [0]{.\EOS\space}%
\providecommand \EOS [0]{\spacefactor3000\relax}%
\providecommand \BibitemShut  [1]{\csname bibitem#1\endcsname}%
\let\auto@bib@innerbib\@empty
\bibitem [{\citenamefont {{Di Valentino}}\ \emph {et~al.}(2022)\citenamefont {{Di Valentino}}, \citenamefont {{Saridakis}},\ and\ \citenamefont {{Riess}}}]{2022NatAs...6.1353D}%
  \BibitemOpen
  \bibfield  {author} {\bibinfo {author} {\bibfnamefont {E.}~\bibnamefont {{Di Valentino}}}, \bibinfo {author} {\bibfnamefont {E.}~\bibnamefont {{Saridakis}}},\ and\ \bibinfo {author} {\bibfnamefont {A.}~\bibnamefont {{Riess}}},\ }\bibfield  {title} {\bibinfo {title} {{Cosmological tensions in the birthplace of the heliocentric model}},\ }\href {https://doi.org/10.1038/s41550-022-01852-3} {\bibfield  {journal} {\bibinfo  {journal} {NatAs}\ }\textbf {\bibinfo {volume} {6}},\ \bibinfo {pages} {1353} (\bibinfo {year} {2022})},\ \Eprint {https://arxiv.org/abs/2211.05248} {arXiv:2211.05248 [astro-ph.CO]} \BibitemShut {NoStop}%
\bibitem [{\citenamefont {Schutz}(1986)}]{Schutz:1986gp}%
  \BibitemOpen
  \bibfield  {author} {\bibinfo {author} {\bibfnamefont {B.~F.}\ \bibnamefont {Schutz}},\ }\bibfield  {title} {\bibinfo {title} {{Determining the Hubble Constant from Gravitational Wave Observations}},\ }\href {https://doi.org/10.1038/323310a0} {\bibfield  {journal} {\bibinfo  {journal} {Nature}\ }\textbf {\bibinfo {volume} {323}},\ \bibinfo {pages} {310} (\bibinfo {year} {1986})}\BibitemShut {NoStop}%
\bibitem [{\citenamefont {{Holz}}\ and\ \citenamefont {{Hughes}}(2005)}]{2005ApJ...629...15H}%
  \BibitemOpen
  \bibfield  {author} {\bibinfo {author} {\bibfnamefont {D.~E.}\ \bibnamefont {{Holz}}}\ and\ \bibinfo {author} {\bibfnamefont {S.~A.}\ \bibnamefont {{Hughes}}},\ }\bibfield  {title} {\bibinfo {title} {{Using Gravitational-Wave Standard Sirens}},\ }\href {https://doi.org/10.1086/431341} {\bibfield  {journal} {\bibinfo  {journal} {ApJ}\ }\textbf {\bibinfo {volume} {629}},\ \bibinfo {pages} {15} (\bibinfo {year} {2005})},\ \Eprint {https://arxiv.org/abs/astro-ph/0504616} {arXiv:astro-ph/0504616 [astro-ph]} \BibitemShut {NoStop}%
\bibitem [{\citenamefont {Abbott}\ \emph {et~al.}(2017)\citenamefont {Abbott} \emph {et~al.}}]{Abbott_2017}%
  \BibitemOpen
  \bibfield  {author} {\bibinfo {author} {\bibfnamefont {B.~P.}\ \bibnamefont {Abbott}} \emph {et~al.},\ }\bibfield  {title} {\bibinfo {title} {Gravitational waves and gamma-rays from a binary neutron star merger: {GW}170817 and {GRB} 170817a},\ }\href {https://doi.org/10.3847/2041-8213/aa920c} {\bibfield  {journal} {\bibinfo  {journal} {The Astrophysical Journal}\ }\textbf {\bibinfo {volume} {848}},\ \bibinfo {pages} {L13} (\bibinfo {year} {2017})}\BibitemShut {NoStop}%
\bibitem [{\citenamefont {{Abbott}}\ \emph {et~al.}(2017)\citenamefont {{Abbott}} \emph {et~al.}}]{2017ApJ...848L..12A}%
  \BibitemOpen
  \bibfield  {author} {\bibinfo {author} {\bibfnamefont {B.~P.}\ \bibnamefont {{Abbott}}} \emph {et~al.},\ }\bibfield  {title} {\bibinfo {title} {{Multi-messenger Observations of a Binary Neutron Star Merger}},\ }\href {https://doi.org/10.3847/2041-8213/aa91c9} {\bibfield  {journal} {\bibinfo  {journal} {ApJL}\ }\textbf {\bibinfo {volume} {848}},\ \bibinfo {eid} {L12} (\bibinfo {year} {2017})},\ \Eprint {https://arxiv.org/abs/1710.05833} {arXiv:1710.05833 [astro-ph.HE]} \BibitemShut {NoStop}%
\bibitem [{\citenamefont {Abbott}\ \emph {et~al.}(2017)\citenamefont {Abbott} \emph {et~al.}}]{LIGOScientific:2017adf}%
  \BibitemOpen
  \bibfield  {author} {\bibinfo {author} {\bibfnamefont {B.~P.}\ \bibnamefont {Abbott}} \emph {et~al.} (\bibinfo {collaboration} {LIGO Scientific, Virgo, 1M2H, Dark Energy Camera GW-E, DES, DLT40, Las Cumbres Observatory, VINROUGE, MASTER}),\ }\bibfield  {title} {\bibinfo {title} {{A gravitational-wave standard siren measurement of the Hubble constant}},\ }\href {https://doi.org/10.1038/nature24471} {\bibfield  {journal} {\bibinfo  {journal} {Nature}\ }\textbf {\bibinfo {volume} {551}},\ \bibinfo {pages} {85} (\bibinfo {year} {2017})},\ \Eprint {https://arxiv.org/abs/1710.05835} {arXiv:1710.05835 [astro-ph.CO]} \BibitemShut {NoStop}%
\bibitem [{\citenamefont {{Abbott}}\ \emph {et~al.}(2019)\citenamefont {{Abbott}} \emph {et~al.}}]{2019PhRvX...9a1001A}%
  \BibitemOpen
  \bibfield  {author} {\bibinfo {author} {\bibfnamefont {B.~P.}\ \bibnamefont {{Abbott}}} \emph {et~al.},\ }\bibfield  {title} {\bibinfo {title} {{Properties of the Binary Neutron Star Merger GW170817}},\ }\href {https://doi.org/10.1103/PhysRevX.9.011001} {\bibfield  {journal} {\bibinfo  {journal} {PhRvX}\ }\textbf {\bibinfo {volume} {9}},\ \bibinfo {eid} {011001} (\bibinfo {year} {2019})},\ \Eprint {https://arxiv.org/abs/1805.11579} {arXiv:1805.11579 [gr-qc]} \BibitemShut {NoStop}%
\bibitem [{\citenamefont {{Chassande-Mottin}}\ \emph {et~al.}(2019)\citenamefont {{Chassande-Mottin}}, \citenamefont {{Leyde}}, \citenamefont {{Mastrogiovanni}},\ and\ \citenamefont {{Steer}}}]{2019PhRvD.100h3514C}%
  \BibitemOpen
  \bibfield  {author} {\bibinfo {author} {\bibfnamefont {E.}~\bibnamefont {{Chassande-Mottin}}}, \bibinfo {author} {\bibfnamefont {K.}~\bibnamefont {{Leyde}}}, \bibinfo {author} {\bibfnamefont {S.}~\bibnamefont {{Mastrogiovanni}}},\ and\ \bibinfo {author} {\bibfnamefont {D.~A.}\ \bibnamefont {{Steer}}},\ }\bibfield  {title} {\bibinfo {title} {{Gravitational wave observations, distance measurement uncertainties, and cosmology}},\ }\href {https://doi.org/10.1103/PhysRevD.100.083514} {\bibfield  {journal} {\bibinfo  {journal} {PhRvD}\ }\textbf {\bibinfo {volume} {100}},\ \bibinfo {eid} {083514} (\bibinfo {year} {2019})},\ \Eprint {https://arxiv.org/abs/1906.02670} {arXiv:1906.02670 [astro-ph.CO]} \BibitemShut {NoStop}%
\bibitem [{\citenamefont {Abbott}\ \emph {et~al.}(2021{\natexlab{a}})\citenamefont {Abbott} \emph {et~al.}}]{LIGOScientific:2020ibl}%
  \BibitemOpen
  \bibfield  {author} {\bibinfo {author} {\bibfnamefont {R.}~\bibnamefont {Abbott}} \emph {et~al.} (\bibinfo {collaboration} {LIGO Scientific, Virgo}),\ }\bibfield  {title} {\bibinfo {title} {{GWTC-2: Compact Binary Coalescences Observed by LIGO and Virgo During the First Half of the Third Observing Run}},\ }\href {https://doi.org/10.1103/PhysRevX.11.021053} {\bibfield  {journal} {\bibinfo  {journal} {Phys. Rev. X}\ }\textbf {\bibinfo {volume} {11}},\ \bibinfo {pages} {021053} (\bibinfo {year} {2021}{\natexlab{a}})},\ \Eprint {https://arxiv.org/abs/2010.14527} {arXiv:2010.14527 [gr-qc]} \BibitemShut {NoStop}%
\bibitem [{\citenamefont {Abbott}\ \emph {et~al.}(2024)\citenamefont {Abbott} \emph {et~al.}}]{LIGOScientific:2021usb}%
  \BibitemOpen
  \bibfield  {author} {\bibinfo {author} {\bibfnamefont {R.}~\bibnamefont {Abbott}} \emph {et~al.} (\bibinfo {collaboration} {LIGO Scientific, VIRGO}),\ }\bibfield  {title} {\bibinfo {title} {{GWTC-2.1: Deep Extended Catalog of Compact Binary Coalescences Observed by LIGO and Virgo During the First Half of the Third Observing Run}},\ }\href {https://doi.org/10.1103/PhysRevD.109.022001} {\bibfield  {journal} {\bibinfo  {journal} {Phys. Rev. D}\ }\textbf {\bibinfo {volume} {109}},\ \bibinfo {pages} {022001} (\bibinfo {year} {2024})},\ \Eprint {https://arxiv.org/abs/2108.01045} {arXiv:2108.01045 [gr-qc]} \BibitemShut {NoStop}%
\bibitem [{\citenamefont {Abbott}\ \emph {et~al.}(2023{\natexlab{a}})\citenamefont {Abbott} \emph {et~al.}}]{KAGRA:2021duu}%
  \BibitemOpen
  \bibfield  {author} {\bibinfo {author} {\bibfnamefont {R.}~\bibnamefont {Abbott}} \emph {et~al.} (\bibinfo {collaboration} {KAGRA, VIRGO, LIGO Scientific}),\ }\bibfield  {title} {\bibinfo {title} {{Population of Merging Compact Binaries Inferred Using Gravitational Waves through GWTC-3}},\ }\href {https://doi.org/10.1103/PhysRevX.13.011048} {\bibfield  {journal} {\bibinfo  {journal} {Phys. Rev. X}\ }\textbf {\bibinfo {volume} {13}},\ \bibinfo {pages} {011048} (\bibinfo {year} {2023}{\natexlab{a}})},\ \Eprint {https://arxiv.org/abs/2111.03634} {arXiv:2111.03634 [astro-ph.HE]} \BibitemShut {NoStop}%
\bibitem [{\citenamefont {Taylor}\ \emph {et~al.}(2012)\citenamefont {Taylor}, \citenamefont {Gair},\ and\ \citenamefont {Mandel}}]{Taylor:2011fs}%
  \BibitemOpen
  \bibfield  {author} {\bibinfo {author} {\bibfnamefont {S.~R.}\ \bibnamefont {Taylor}}, \bibinfo {author} {\bibfnamefont {J.~R.}\ \bibnamefont {Gair}},\ and\ \bibinfo {author} {\bibfnamefont {I.}~\bibnamefont {Mandel}},\ }\bibfield  {title} {\bibinfo {title} {{Hubble without the Hubble: Cosmology using advanced gravitational-wave detectors alone}},\ }\href {https://doi.org/10.1103/PhysRevD.85.023535} {\bibfield  {journal} {\bibinfo  {journal} {Phys. Rev. D}\ }\textbf {\bibinfo {volume} {85}},\ \bibinfo {pages} {023535} (\bibinfo {year} {2012})},\ \Eprint {https://arxiv.org/abs/1108.5161} {arXiv:1108.5161 [gr-qc]} \BibitemShut {NoStop}%
\bibitem [{\citenamefont {Wysocki}\ \emph {et~al.}(2019)\citenamefont {Wysocki}, \citenamefont {Lange},\ and\ \citenamefont {O'Shaughnessy}}]{Wysocki:2018mpo}%
  \BibitemOpen
  \bibfield  {author} {\bibinfo {author} {\bibfnamefont {D.}~\bibnamefont {Wysocki}}, \bibinfo {author} {\bibfnamefont {J.}~\bibnamefont {Lange}},\ and\ \bibinfo {author} {\bibfnamefont {R.}~\bibnamefont {O'Shaughnessy}},\ }\bibfield  {title} {\bibinfo {title} {{Reconstructing phenomenological distributions of compact binaries via gravitational wave observations}},\ }\href {https://doi.org/10.1103/PhysRevD.100.043012} {\bibfield  {journal} {\bibinfo  {journal} {Phys. Rev. D}\ }\textbf {\bibinfo {volume} {100}},\ \bibinfo {pages} {043012} (\bibinfo {year} {2019})},\ \Eprint {https://arxiv.org/abs/1805.06442} {arXiv:1805.06442 [gr-qc]} \BibitemShut {NoStop}%
\bibitem [{\citenamefont {Farr}\ \emph {et~al.}(2019)\citenamefont {Farr}, \citenamefont {Fishbach}, \citenamefont {Ye},\ and\ \citenamefont {Holz}}]{Farr:2019twy}%
  \BibitemOpen
  \bibfield  {author} {\bibinfo {author} {\bibfnamefont {W.~M.}\ \bibnamefont {Farr}}, \bibinfo {author} {\bibfnamefont {M.}~\bibnamefont {Fishbach}}, \bibinfo {author} {\bibfnamefont {J.}~\bibnamefont {Ye}},\ and\ \bibinfo {author} {\bibfnamefont {D.}~\bibnamefont {Holz}},\ }\bibfield  {title} {\bibinfo {title} {{A Future Percent-Level Measurement of the Hubble Expansion at Redshift 0.8 With Advanced LIGO}},\ }\href {https://doi.org/10.3847/2041-8213/ab4284} {\bibfield  {journal} {\bibinfo  {journal} {Astrophys. J. Lett.}\ }\textbf {\bibinfo {volume} {883}},\ \bibinfo {pages} {L42} (\bibinfo {year} {2019})},\ \Eprint {https://arxiv.org/abs/1908.09084} {arXiv:1908.09084 [astro-ph.CO]} \BibitemShut {NoStop}%
\bibitem [{\citenamefont {Mastrogiovanni}\ \emph {et~al.}(2021)\citenamefont {Mastrogiovanni}, \citenamefont {Leyde}, \citenamefont {Karathanasis}, \citenamefont {Chassande-Mottin}, \citenamefont {Steer}, \citenamefont {Gair}, \citenamefont {Ghosh}, \citenamefont {Gray}, \citenamefont {Mukherjee},\ and\ \citenamefont {Rinaldi}}]{Mastrogiovanni:2021wsd}%
  \BibitemOpen
  \bibfield  {author} {\bibinfo {author} {\bibfnamefont {S.}~\bibnamefont {Mastrogiovanni}}, \bibinfo {author} {\bibfnamefont {K.}~\bibnamefont {Leyde}}, \bibinfo {author} {\bibfnamefont {C.}~\bibnamefont {Karathanasis}}, \bibinfo {author} {\bibfnamefont {E.}~\bibnamefont {Chassande-Mottin}}, \bibinfo {author} {\bibfnamefont {D.~A.}\ \bibnamefont {Steer}}, \bibinfo {author} {\bibfnamefont {J.}~\bibnamefont {Gair}}, \bibinfo {author} {\bibfnamefont {A.}~\bibnamefont {Ghosh}}, \bibinfo {author} {\bibfnamefont {R.}~\bibnamefont {Gray}}, \bibinfo {author} {\bibfnamefont {S.}~\bibnamefont {Mukherjee}},\ and\ \bibinfo {author} {\bibfnamefont {S.}~\bibnamefont {Rinaldi}},\ }\bibfield  {title} {\bibinfo {title} {{On the importance of source population models for gravitational-wave cosmology}},\ }\href {https://doi.org/10.1103/PhysRevD.104.062009} {\bibfield  {journal} {\bibinfo  {journal} {Phys. Rev. D}\ }\textbf {\bibinfo {volume} {104}},\ \bibinfo {pages} {062009} (\bibinfo {year} {2021})},\ \Eprint
  {https://arxiv.org/abs/2103.14663} {arXiv:2103.14663 [gr-qc]} \BibitemShut {NoStop}%
\bibitem [{\citenamefont {Mancarella}\ \emph {et~al.}(2022)\citenamefont {Mancarella}, \citenamefont {Genoud-Prachex},\ and\ \citenamefont {Maggiore}}]{Mancarella:2021ecn}%
  \BibitemOpen
  \bibfield  {author} {\bibinfo {author} {\bibfnamefont {M.}~\bibnamefont {Mancarella}}, \bibinfo {author} {\bibfnamefont {E.}~\bibnamefont {Genoud-Prachex}},\ and\ \bibinfo {author} {\bibfnamefont {M.}~\bibnamefont {Maggiore}},\ }\bibfield  {title} {\bibinfo {title} {{Cosmology and modified gravitational wave propagation from binary black hole population models}},\ }\href {https://doi.org/10.1103/PhysRevD.105.064030} {\bibfield  {journal} {\bibinfo  {journal} {Phys. Rev. D}\ }\textbf {\bibinfo {volume} {105}},\ \bibinfo {pages} {064030} (\bibinfo {year} {2022})},\ \Eprint {https://arxiv.org/abs/2112.05728} {arXiv:2112.05728 [gr-qc]} \BibitemShut {NoStop}%
\bibitem [{\citenamefont {Mukherjee}(2022)}]{Mukherjee:2021rtw}%
  \BibitemOpen
  \bibfield  {author} {\bibinfo {author} {\bibfnamefont {S.}~\bibnamefont {Mukherjee}},\ }\bibfield  {title} {\bibinfo {title} {{The redshift dependence of black hole mass distribution: is it reliable for standard sirens cosmology?}},\ }\href {https://doi.org/10.1093/mnras/stac2152} {\bibfield  {journal} {\bibinfo  {journal} {Mon. Not. Roy. Astron. Soc.}\ }\textbf {\bibinfo {volume} {515}},\ \bibinfo {pages} {5495} (\bibinfo {year} {2022})},\ \Eprint {https://arxiv.org/abs/2112.10256} {arXiv:2112.10256 [astro-ph.CO]} \BibitemShut {NoStop}%
\bibitem [{\citenamefont {Gray}(2021)}]{Gray:2021qfw}%
  \BibitemOpen
  \bibfield  {author} {\bibinfo {author} {\bibfnamefont {R.}~\bibnamefont {Gray}},\ }\emph {\bibinfo {title} {{Gravitational wave cosmology: measuring the Hubble constant with dark standard sirens}}},\ \href {https://doi.org/10.5525/gla.thesis.82438} {Ph.D. thesis},\ \bibinfo  {school} {University of Glasgow} (\bibinfo {year} {2021})\BibitemShut {NoStop}%
\bibitem [{\citenamefont {Gray}\ \emph {et~al.}(2022)\citenamefont {Gray}, \citenamefont {Messenger},\ and\ \citenamefont {Veitch}}]{Gray:2021sew}%
  \BibitemOpen
  \bibfield  {author} {\bibinfo {author} {\bibfnamefont {R.}~\bibnamefont {Gray}}, \bibinfo {author} {\bibfnamefont {C.}~\bibnamefont {Messenger}},\ and\ \bibinfo {author} {\bibfnamefont {J.}~\bibnamefont {Veitch}},\ }\bibfield  {title} {\bibinfo {title} {{A pixelated approach to galaxy catalogue incompleteness: improving the dark siren measurement of the Hubble constant}},\ }\href {https://doi.org/10.1093/mnras/stac366} {\bibfield  {journal} {\bibinfo  {journal} {Mon. Not. Roy. Astron. Soc.}\ }\textbf {\bibinfo {volume} {512}},\ \bibinfo {pages} {1127} (\bibinfo {year} {2022})},\ \Eprint {https://arxiv.org/abs/2111.04629} {arXiv:2111.04629 [astro-ph.CO]} \BibitemShut {NoStop}%
\bibitem [{\citenamefont {Gair}\ \emph {et~al.}(2023{\natexlab{a}})\citenamefont {Gair} \emph {et~al.}}]{Gair:2022zsa}%
  \BibitemOpen
  \bibfield  {author} {\bibinfo {author} {\bibfnamefont {J.~R.}\ \bibnamefont {Gair}} \emph {et~al.},\ }\bibfield  {title} {\bibinfo {title} {{The Hitchhiker's guide to the galaxy catalog approach for gravitational wave cosmology}},\ }\href {https://doi.org/10.3847/1538-3881/acca78} {\bibfield  {journal} {\bibinfo  {journal} {The Astronomical Journal}\ }\textbf {\bibinfo {volume} {166}},\ \bibinfo {pages} {22} (\bibinfo {year} {2023}{\natexlab{a}})},\ \Eprint {https://arxiv.org/abs/2212.08694} {arXiv:2212.08694 [gr-qc]} \BibitemShut {NoStop}%
\bibitem [{\citenamefont {Leyde}\ \emph {et~al.}(2022)\citenamefont {Leyde}, \citenamefont {Mastrogiovanni}, \citenamefont {Steer}, \citenamefont {Chassande-Mottin},\ and\ \citenamefont {Karathanasis}}]{Leyde:2022orh}%
  \BibitemOpen
  \bibfield  {author} {\bibinfo {author} {\bibfnamefont {K.}~\bibnamefont {Leyde}}, \bibinfo {author} {\bibfnamefont {S.}~\bibnamefont {Mastrogiovanni}}, \bibinfo {author} {\bibfnamefont {D.~A.}\ \bibnamefont {Steer}}, \bibinfo {author} {\bibfnamefont {E.}~\bibnamefont {Chassande-Mottin}},\ and\ \bibinfo {author} {\bibfnamefont {C.}~\bibnamefont {Karathanasis}},\ }\bibfield  {title} {\bibinfo {title} {{Current and future constraints on cosmology and modified gravitational wave friction from binary black holes}},\ }\href {https://doi.org/10.1088/1475-7516/2022/09/012} {\bibfield  {journal} {\bibinfo  {journal} {JCAP}\ }\textbf {\bibinfo {volume} {09}},\ \bibinfo {pages} {012}},\ \Eprint {https://arxiv.org/abs/2202.00025} {arXiv:2202.00025 [gr-qc]} \BibitemShut {NoStop}%
\bibitem [{\citenamefont {Ezquiaga}\ and\ \citenamefont {Holz}(2022)}]{Ezquiaga:2022zkx}%
  \BibitemOpen
  \bibfield  {author} {\bibinfo {author} {\bibfnamefont {J.~M.}\ \bibnamefont {Ezquiaga}}\ and\ \bibinfo {author} {\bibfnamefont {D.~E.}\ \bibnamefont {Holz}},\ }\bibfield  {title} {\bibinfo {title} {{Spectral Sirens: Cosmology from the Full Mass Distribution of Compact Binaries}},\ }\href {https://doi.org/10.1103/PhysRevLett.129.061102} {\bibfield  {journal} {\bibinfo  {journal} {Phys. Rev. Lett.}\ }\textbf {\bibinfo {volume} {129}},\ \bibinfo {pages} {061102} (\bibinfo {year} {2022})},\ \Eprint {https://arxiv.org/abs/2202.08240} {arXiv:2202.08240 [astro-ph.CO]} \BibitemShut {NoStop}%
\bibitem [{\citenamefont {{Borghi}}\ \emph {et~al.}(2024)\citenamefont {{Borghi}}, \citenamefont {{Mancarella}}, \citenamefont {{Moresco}}, \citenamefont {{Tagliazucchi}}, \citenamefont {{Iacovelli}}, \citenamefont {{Cimatti}},\ and\ \citenamefont {{Maggiore}}}]{2023arXiv231205302B}%
  \BibitemOpen
  \bibfield  {author} {\bibinfo {author} {\bibfnamefont {N.}~\bibnamefont {{Borghi}}}, \bibinfo {author} {\bibfnamefont {M.}~\bibnamefont {{Mancarella}}}, \bibinfo {author} {\bibfnamefont {M.}~\bibnamefont {{Moresco}}}, \bibinfo {author} {\bibfnamefont {M.}~\bibnamefont {{Tagliazucchi}}}, \bibinfo {author} {\bibfnamefont {F.}~\bibnamefont {{Iacovelli}}}, \bibinfo {author} {\bibfnamefont {A.}~\bibnamefont {{Cimatti}}},\ and\ \bibinfo {author} {\bibfnamefont {M.}~\bibnamefont {{Maggiore}}},\ }\bibfield  {title} {\bibinfo {title} {{Cosmology and Astrophysics with Standard Sirens and Galaxy Catalogs in View of Future Gravitational Wave Observations}},\ }\href {https://doi.org/10.3847/1538-4357/ad20eb} {\bibfield  {journal} {\bibinfo  {journal} {The Astrophysical Journal}\ }\textbf {\bibinfo {volume} {964}},\ \bibinfo {pages} {191} (\bibinfo {year} {2024})},\ \Eprint {https://arxiv.org/abs/2312.05302} {arXiv:2312.05302 [astro-ph.CO]} \BibitemShut {NoStop}%
\bibitem [{\citenamefont {{Karathanasis}}\ \emph {et~al.}(2023)\citenamefont {{Karathanasis}}, \citenamefont {{Mukherjee}},\ and\ \citenamefont {{Mastrogiovanni}}}]{2023MNRAS.523.4539K}%
  \BibitemOpen
  \bibfield  {author} {\bibinfo {author} {\bibfnamefont {C.}~\bibnamefont {{Karathanasis}}}, \bibinfo {author} {\bibfnamefont {S.}~\bibnamefont {{Mukherjee}}},\ and\ \bibinfo {author} {\bibfnamefont {S.}~\bibnamefont {{Mastrogiovanni}}},\ }\bibfield  {title} {\bibinfo {title} {{Binary black holes population and cosmology in new lights: signature of PISN mass and formation channel in GWTC-3}},\ }\href {https://doi.org/10.1093/mnras/stad1373} {\bibfield  {journal} {\bibinfo  {journal} {MNRAS}\ }\textbf {\bibinfo {volume} {523}},\ \bibinfo {pages} {4539} (\bibinfo {year} {2023})},\ \Eprint {https://arxiv.org/abs/2204.13495} {arXiv:2204.13495 [astro-ph.CO]} \BibitemShut {NoStop}%
\bibitem [{\citenamefont {{Pierra}}\ \emph {et~al.}(2024)\citenamefont {{Pierra}}, \citenamefont {{Mastrogiovanni}}, \citenamefont {{Perri{\`e}s}},\ and\ \citenamefont {{Mapelli}}}]{2023arXiv231211627P}%
  \BibitemOpen
  \bibfield  {author} {\bibinfo {author} {\bibfnamefont {G.}~\bibnamefont {{Pierra}}}, \bibinfo {author} {\bibfnamefont {S.}~\bibnamefont {{Mastrogiovanni}}}, \bibinfo {author} {\bibfnamefont {S.}~\bibnamefont {{Perri{\`e}s}}},\ and\ \bibinfo {author} {\bibfnamefont {M.}~\bibnamefont {{Mapelli}}},\ }\bibfield  {title} {\bibinfo {title} {{A Study of Systematics on the Cosmological Inference of the Hubble Constant from Gravitational Wave Standard Sirens}},\ }\href {https://doi.org/10.1103/PhysRevD.109.083504} {\bibfield  {journal} {\bibinfo  {journal} {Phys. Rev. D}\ }\textbf {\bibinfo {volume} {109}},\ \bibinfo {pages} {083504} (\bibinfo {year} {2024})},\ \Eprint {https://arxiv.org/abs/2312.11627} {arXiv:2312.11627 [astro-ph.CO]} \BibitemShut {NoStop}%
\bibitem [{\citenamefont {Abbott}\ \emph {et~al.}(2023{\natexlab{b}})\citenamefont {Abbott} \emph {et~al.}}]{LIGOScientific:2021aug}%
  \BibitemOpen
  \bibfield  {author} {\bibinfo {author} {\bibfnamefont {R.}~\bibnamefont {Abbott}} \emph {et~al.} (\bibinfo {collaboration} {LIGO Scientific, Virgo,, KAGRA, VIRGO}),\ }\bibfield  {title} {\bibinfo {title} {{Constraints on the Cosmic Expansion History from GWTC\textendash{}3}},\ }\href {https://doi.org/10.3847/1538-4357/ac74bb} {\bibfield  {journal} {\bibinfo  {journal} {Astrophys. J.}\ }\textbf {\bibinfo {volume} {949}},\ \bibinfo {pages} {76} (\bibinfo {year} {2023}{\natexlab{b}})},\ \Eprint {https://arxiv.org/abs/2111.03604} {arXiv:2111.03604 [astro-ph.CO]} \BibitemShut {NoStop}%
\bibitem [{\citenamefont {{Punturo}}\ \emph {et~al.}(2010)\citenamefont {{Punturo}}, \citenamefont {{Abernathy}}, \citenamefont {{Acernese}}, \citenamefont {{Allen}}, \citenamefont {{Andersson}}, \citenamefont {{Arun}}, \citenamefont {{Barone}}, \citenamefont {{Barr}}, \citenamefont {{Barsuglia}}, \citenamefont {{Beker}} \emph {et~al.}}]{2010CQGra..27s4002P}%
  \BibitemOpen
  \bibfield  {author} {\bibinfo {author} {\bibfnamefont {M.}~\bibnamefont {{Punturo}}}, \bibinfo {author} {\bibfnamefont {M.}~\bibnamefont {{Abernathy}}}, \bibinfo {author} {\bibfnamefont {F.}~\bibnamefont {{Acernese}}}, \bibinfo {author} {\bibfnamefont {B.}~\bibnamefont {{Allen}}}, \bibinfo {author} {\bibfnamefont {N.}~\bibnamefont {{Andersson}}}, \bibinfo {author} {\bibfnamefont {K.}~\bibnamefont {{Arun}}}, \bibinfo {author} {\bibfnamefont {F.}~\bibnamefont {{Barone}}}, \bibinfo {author} {\bibfnamefont {B.}~\bibnamefont {{Barr}}}, \bibinfo {author} {\bibfnamefont {M.}~\bibnamefont {{Barsuglia}}}, \bibinfo {author} {\bibfnamefont {M.}~\bibnamefont {{Beker}}}, \emph {et~al.},\ }\bibfield  {title} {\bibinfo {title} {{The Einstein Telescope: a third-generation gravitational wave observatory}},\ }\href {https://doi.org/10.1088/0264-9381/27/19/194002} {\bibfield  {journal} {\bibinfo  {journal} {CQGra}\ }\textbf {\bibinfo {volume} {27}},\ \bibinfo {eid} {194002} (\bibinfo {year} {2010})}\BibitemShut {NoStop}%
\bibitem [{\citenamefont {Maggiore}\ \emph {et~al.}(2020)\citenamefont {Maggiore} \emph {et~al.}}]{Maggiore:2019uih}%
  \BibitemOpen
  \bibfield  {author} {\bibinfo {author} {\bibfnamefont {M.}~\bibnamefont {Maggiore}} \emph {et~al.},\ }\bibfield  {title} {\bibinfo {title} {{Science Case for the Einstein Telescope}},\ }\href {https://doi.org/10.1088/1475-7516/2020/03/050} {\bibfield  {journal} {\bibinfo  {journal} {JCAP}\ }\textbf {\bibinfo {volume} {03}},\ \bibinfo {pages} {050}},\ \Eprint {https://arxiv.org/abs/1912.02622} {arXiv:1912.02622 [astro-ph.CO]} \BibitemShut {NoStop}%
\bibitem [{\citenamefont {{Branchesi}}\ \emph {et~al.}(2023)\citenamefont {{Branchesi}}, \citenamefont {{Maggiore}}, \citenamefont {{Alonso}}, \citenamefont {{Badger}}, \citenamefont {{Banerjee}}, \citenamefont {{Beirnaert}}, \citenamefont {{Belgacem}}, \citenamefont {{Bhagwat}}, \citenamefont {{Boileau}}, \citenamefont {{Borhanian}} \emph {et~al.}}]{2023JCAP...07..068B}%
  \BibitemOpen
  \bibfield  {author} {\bibinfo {author} {\bibfnamefont {M.}~\bibnamefont {{Branchesi}}}, \bibinfo {author} {\bibfnamefont {M.}~\bibnamefont {{Maggiore}}}, \bibinfo {author} {\bibfnamefont {D.}~\bibnamefont {{Alonso}}}, \bibinfo {author} {\bibfnamefont {C.}~\bibnamefont {{Badger}}}, \bibinfo {author} {\bibfnamefont {B.}~\bibnamefont {{Banerjee}}}, \bibinfo {author} {\bibfnamefont {F.}~\bibnamefont {{Beirnaert}}}, \bibinfo {author} {\bibfnamefont {E.}~\bibnamefont {{Belgacem}}}, \bibinfo {author} {\bibfnamefont {S.}~\bibnamefont {{Bhagwat}}}, \bibinfo {author} {\bibfnamefont {G.}~\bibnamefont {{Boileau}}}, \bibinfo {author} {\bibfnamefont {S.}~\bibnamefont {{Borhanian}}}, \emph {et~al.},\ }\bibfield  {title} {\bibinfo {title} {{Science with the Einstein Telescope: a comparison of different designs}},\ }\href {https://doi.org/10.1088/1475-7516/2023/07/068} {\bibfield  {journal} {\bibinfo  {journal} {JCAP}\ }\textbf {\bibinfo {volume} {2023}}\bibfield  {number} {\bibinfo  {number} { (7)},\ \bibinfo {eid}
  {068}},\ }\Eprint {https://arxiv.org/abs/2303.15923} {arXiv:2303.15923 [gr-qc]} \BibitemShut {NoStop}%
\bibitem [{\citenamefont {{Antonelli}}\ \emph {et~al.}(2009)\citenamefont {{Antonelli}}, \citenamefont {{D'Avanzo}}, \citenamefont {{Perna}}, \citenamefont {{Amati}}, \citenamefont {{Covino}}, \citenamefont {{Cutini}}, \citenamefont {{D'Elia}}, \citenamefont {{Gallozzi}}, \citenamefont {{Grazian}}, \citenamefont {{Palazzi}},\ and\ \citenamefont {et~al.}}]{2009A&A...507L..45A}%
  \BibitemOpen
  \bibfield  {author} {\bibinfo {author} {\bibfnamefont {L.~A.}\ \bibnamefont {{Antonelli}}}, \bibinfo {author} {\bibfnamefont {P.}~\bibnamefont {{D'Avanzo}}}, \bibinfo {author} {\bibfnamefont {R.}~\bibnamefont {{Perna}}}, \bibinfo {author} {\bibfnamefont {L.}~\bibnamefont {{Amati}}}, \bibinfo {author} {\bibfnamefont {S.}~\bibnamefont {{Covino}}}, \bibinfo {author} {\bibfnamefont {S.}~\bibnamefont {{Cutini}}}, \bibinfo {author} {\bibfnamefont {V.}~\bibnamefont {{D'Elia}}}, \bibinfo {author} {\bibfnamefont {S.}~\bibnamefont {{Gallozzi}}}, \bibinfo {author} {\bibfnamefont {A.}~\bibnamefont {{Grazian}}}, \bibinfo {author} {\bibfnamefont {E.}~\bibnamefont {{Palazzi}}},\ and\ \bibinfo {author} {\bibnamefont {et~al.}},\ }\bibfield  {title} {\bibinfo {title} {{GRB 090426: the farthest short gamma-ray burst?}},\ }\href {https://doi.org/10.1051/0004-6361/200913062} {\bibfield  {journal} {\bibinfo  {journal} {A\&A}\ }\textbf {\bibinfo {volume} {507}},\ \bibinfo {pages} {L45} (\bibinfo {year} {2009})},\ \Eprint
  {https://arxiv.org/abs/0911.0046} {arXiv:0911.0046 [astro-ph.HE]} \BibitemShut {NoStop}%
\bibitem [{\citenamefont {Collaboration}\ \emph {et~al.}(2015)\citenamefont {Collaboration} \emph {et~al.}}]{2015}%
  \BibitemOpen
  \bibfield  {author} {\bibinfo {author} {\bibfnamefont {T.~L.~S.}\ \bibnamefont {Collaboration}} \emph {et~al.},\ }\bibfield  {title} {\bibinfo {title} {Advanced ligo},\ }\href {https://doi.org/10.1088/0264-9381/32/7/074001} {\bibfield  {journal} {\bibinfo  {journal} {Classical and Quantum Gravity}\ }\textbf {\bibinfo {volume} {32}},\ \bibinfo {pages} {074001} (\bibinfo {year} {2015})}\BibitemShut {NoStop}%
\bibitem [{\citenamefont {Acernese}\ \emph {et~al.}(2014)\citenamefont {Acernese} \emph {et~al.}}]{Acernese_2014}%
  \BibitemOpen
  \bibfield  {author} {\bibinfo {author} {\bibfnamefont {F.}~\bibnamefont {Acernese}} \emph {et~al.},\ }\bibfield  {title} {\bibinfo {title} {Advanced virgo: a second-generation interferometric gravitational wave detector},\ }\href {https://doi.org/10.1088/0264-9381/32/2/024001} {\bibfield  {journal} {\bibinfo  {journal} {Classical and Quantum Gravity}\ }\textbf {\bibinfo {volume} {32}},\ \bibinfo {pages} {024001} (\bibinfo {year} {2014})}\BibitemShut {NoStop}%
\bibitem [{\citenamefont {Akutsu}\ \emph {et~al.}(2020)\citenamefont {Akutsu} \emph {et~al.}}]{10.1093/ptep/ptaa125}%
  \BibitemOpen
  \bibfield  {author} {\bibinfo {author} {\bibfnamefont {T.}~\bibnamefont {Akutsu}} \emph {et~al.},\ }\bibfield  {title} {\bibinfo {title} {{Overview of KAGRA: Detector design and construction history}},\ }\href {https://doi.org/10.1093/ptep/ptaa125} {\bibfield  {journal} {\bibinfo  {journal} {Progress of Theoretical and Experimental Physics}\ }\textbf {\bibinfo {volume} {2021}},\ \bibinfo {pages} {05A101} (\bibinfo {year} {2020})},\ \Eprint {https://arxiv.org/abs/https://academic.oup.com/ptep/article-pdf/2021/5/05A101/37974994/ptaa125.pdf} {https://academic.oup.com/ptep/article-pdf/2021/5/05A101/37974994/ptaa125.pdf} \BibitemShut {NoStop}%
\bibitem [{\citenamefont {Abbott}\ \emph {et~al.}(2018)\citenamefont {Abbott} \emph {et~al.}}]{KAGRA:2013rdx}%
  \BibitemOpen
  \bibfield  {author} {\bibinfo {author} {\bibfnamefont {B.~P.}\ \bibnamefont {Abbott}} \emph {et~al.} (\bibinfo {collaboration} {KAGRA, LIGO Scientific, VIRGO}),\ }\bibfield  {title} {\bibinfo {title} {{Prospects for observing and localizing gravitational-wave transients with Advanced LIGO, Advanced Virgo and KAGRA}},\ }\href {https://doi.org/10.1007/s41114-020-00026-9} {\bibfield  {journal} {\bibinfo  {journal} {Living Rev. Rel.}\ }\textbf {\bibinfo {volume} {21}},\ \bibinfo {pages} {3} (\bibinfo {year} {2018})},\ \Eprint {https://arxiv.org/abs/1304.0670} {arXiv:1304.0670 [gr-qc]} \BibitemShut {NoStop}%
\bibitem [{\citenamefont {{Mastrogiovanni}}\ \emph {et~al.}(2020)\citenamefont {{Mastrogiovanni}}, \citenamefont {{Steer}},\ and\ \citenamefont {{Barsuglia}}}]{2020PhRvD.102d4009M}%
  \BibitemOpen
  \bibfield  {author} {\bibinfo {author} {\bibfnamefont {S.}~\bibnamefont {{Mastrogiovanni}}}, \bibinfo {author} {\bibfnamefont {D.~A.}\ \bibnamefont {{Steer}}},\ and\ \bibinfo {author} {\bibfnamefont {M.}~\bibnamefont {{Barsuglia}}},\ }\bibfield  {title} {\bibinfo {title} {{Probing modified gravity theories and cosmology using gravitational-waves and associated electromagnetic counterparts}},\ }\href {https://doi.org/10.1103/PhysRevD.102.044009} {\bibfield  {journal} {\bibinfo  {journal} {PhRvD}\ }\textbf {\bibinfo {volume} {102}},\ \bibinfo {eid} {044009} (\bibinfo {year} {2020})},\ \Eprint {https://arxiv.org/abs/2004.01632} {arXiv:2004.01632 [gr-qc]} \BibitemShut {NoStop}%
\bibitem [{\citenamefont {{Enea Romano}}\ and\ \citenamefont {{Sakellariadou}}(2023)}]{Romano:2023bge}%
  \BibitemOpen
  \bibfield  {author} {\bibinfo {author} {\bibfnamefont {A.}~\bibnamefont {{Enea Romano}}}\ and\ \bibinfo {author} {\bibfnamefont {M.}~\bibnamefont {{Sakellariadou}}},\ }\bibfield  {title} {\bibinfo {title} {{Constraining the time evolution of the propagation speed of gravitational waves with multimessenger astronomy}},\ }\href {https://doi.org/10.48550/arXiv.2309.10903} {\bibfield  {journal} {\bibinfo  {journal} {arXiv}\ ,\ \bibinfo {eid} {arXiv:2309.10903}} (\bibinfo {year} {2023})},\ \Eprint {https://arxiv.org/abs/2309.10903} {arXiv:2309.10903 [gr-qc]} \BibitemShut {NoStop}%
\bibitem [{\citenamefont {{Romano}}(2024)}]{2024PhLB..85138572R}%
  \BibitemOpen
  \bibfield  {author} {\bibinfo {author} {\bibfnamefont {A.~E.}\ \bibnamefont {{Romano}}},\ }\bibfield  {title} {\bibinfo {title} {{Effective speed of gravitational waves}},\ }\href {https://doi.org/10.1016/j.physletb.2024.138572} {\bibfield  {journal} {\bibinfo  {journal} {PhLB}\ }\textbf {\bibinfo {volume} {851}},\ \bibinfo {eid} {138572} (\bibinfo {year} {2024})},\ \Eprint {https://arxiv.org/abs/2211.05760} {arXiv:2211.05760 [gr-qc]} \BibitemShut {NoStop}%
\bibitem [{\citenamefont {Ghosh}\ \emph {et~al.}(2023)\citenamefont {Ghosh}, \citenamefont {Nair}, \citenamefont {Pathak}, \citenamefont {Sarkar},\ and\ \citenamefont {Sengupta}}]{PhysRevD.108.124017}%
  \BibitemOpen
  \bibfield  {author} {\bibinfo {author} {\bibfnamefont {R.}~\bibnamefont {Ghosh}}, \bibinfo {author} {\bibfnamefont {S.}~\bibnamefont {Nair}}, \bibinfo {author} {\bibfnamefont {L.}~\bibnamefont {Pathak}}, \bibinfo {author} {\bibfnamefont {S.}~\bibnamefont {Sarkar}},\ and\ \bibinfo {author} {\bibfnamefont {A.~S.}\ \bibnamefont {Sengupta}},\ }\bibfield  {title} {\bibinfo {title} {Does the speed of gravitational waves depend on the source velocity?},\ }\href {https://doi.org/10.1103/PhysRevD.108.124017} {\bibfield  {journal} {\bibinfo  {journal} {Phys. Rev. D}\ }\textbf {\bibinfo {volume} {108}},\ \bibinfo {pages} {124017} (\bibinfo {year} {2023})}\BibitemShut {NoStop}%
\bibitem [{\citenamefont {Mandel}\ \emph {et~al.}(2019)\citenamefont {Mandel}, \citenamefont {Farr},\ and\ \citenamefont {Gair}}]{Mandel_2019}%
  \BibitemOpen
  \bibfield  {author} {\bibinfo {author} {\bibfnamefont {I.}~\bibnamefont {Mandel}}, \bibinfo {author} {\bibfnamefont {W.~M.}\ \bibnamefont {Farr}},\ and\ \bibinfo {author} {\bibfnamefont {J.~R.}\ \bibnamefont {Gair}},\ }\bibfield  {title} {\bibinfo {title} {Extracting distribution parameters from multiple uncertain observations with selection biases},\ }\href {https://doi.org/10.1093/mnras/stz896} {\bibfield  {journal} {\bibinfo  {journal} {Monthly Notices of the Royal Astronomical Society}\ }\textbf {\bibinfo {volume} {486}},\ \bibinfo {pages} {1086} (\bibinfo {year} {2019})}\BibitemShut {NoStop}%
\bibitem [{\citenamefont {Mastrogiovanni}\ \emph {et~al.}(2024)\citenamefont {Mastrogiovanni}, \citenamefont {Pierra}, \citenamefont {Perri\`es}, \citenamefont {Laghi}, \citenamefont {Caneva~Santoro}, \citenamefont {Ghosh}, \citenamefont {Gray}, \citenamefont {Karathanasis},\ and\ \citenamefont {Leyde}}]{mastrogiovanni2023icarogw}%
  \BibitemOpen
  \bibfield  {author} {\bibinfo {author} {\bibfnamefont {S.}~\bibnamefont {Mastrogiovanni}}, \bibinfo {author} {\bibfnamefont {G.}~\bibnamefont {Pierra}}, \bibinfo {author} {\bibfnamefont {S.}~\bibnamefont {Perri\`es}}, \bibinfo {author} {\bibfnamefont {D.}~\bibnamefont {Laghi}}, \bibinfo {author} {\bibfnamefont {G.}~\bibnamefont {Caneva~Santoro}}, \bibinfo {author} {\bibfnamefont {A.}~\bibnamefont {Ghosh}}, \bibinfo {author} {\bibfnamefont {R.}~\bibnamefont {Gray}}, \bibinfo {author} {\bibfnamefont {C.}~\bibnamefont {Karathanasis}},\ and\ \bibinfo {author} {\bibfnamefont {K.}~\bibnamefont {Leyde}},\ }\bibfield  {title} {\bibinfo {title} {{ICAROGW: A python package for inference of astrophysical population properties of noisy, heterogeneous, and incomplete observations}},\ }\href {https://doi.org/10.1051/0004-6361/202347007} {\bibfield  {journal} {\bibinfo  {journal} {Astron. Astrophys.}\ }\textbf {\bibinfo {volume} {682}},\ \bibinfo {pages} {A167} (\bibinfo {year} {2024})},\ \Eprint
  {https://arxiv.org/abs/2305.17973} {arXiv:2305.17973 [astro-ph.CO]} \BibitemShut {NoStop}%
\bibitem [{\citenamefont {Aghanim}\ \emph {et~al.}(2020)\citenamefont {Aghanim} \emph {et~al.}}]{2020}%
  \BibitemOpen
  \bibfield  {author} {\bibinfo {author} {\bibfnamefont {N.}~\bibnamefont {Aghanim}} \emph {et~al.} (\bibinfo {collaboration} {Planck}),\ }\bibfield  {title} {\bibinfo {title} {{Planck 2018 results. VI. Cosmological parameters}},\ }\href {https://doi.org/10.1051/0004-6361/201833910} {\bibfield  {journal} {\bibinfo  {journal} {Astron. Astrophys.}\ }\textbf {\bibinfo {volume} {641}},\ \bibinfo {pages} {A6} (\bibinfo {year} {2020})},\ \bibinfo {note} {[Erratum: Astron.Astrophys. 652, C4 (2021)]},\ \Eprint {https://arxiv.org/abs/1807.06209} {arXiv:1807.06209 [astro-ph.CO]} \BibitemShut {NoStop}%
\bibitem [{\citenamefont {Abbott}\ \emph {et~al.}(2023{\natexlab{c}})\citenamefont {Abbott} \emph {et~al.}}]{theligoscientificcollaboration2022population}%
  \BibitemOpen
  \bibfield  {author} {\bibinfo {author} {\bibfnamefont {R.}~\bibnamefont {Abbott}} \emph {et~al.} (\bibinfo {collaboration} {KAGRA, VIRGO, LIGO Scientific}),\ }\bibfield  {title} {\bibinfo {title} {{Population of Merging Compact Binaries Inferred Using Gravitational Waves through GWTC-3}},\ }\href {https://doi.org/10.1103/PhysRevX.13.011048} {\bibfield  {journal} {\bibinfo  {journal} {Phys. Rev. X}\ }\textbf {\bibinfo {volume} {13}},\ \bibinfo {pages} {011048} (\bibinfo {year} {2023}{\natexlab{c}})},\ \Eprint {https://arxiv.org/abs/2111.03634} {arXiv:2111.03634 [astro-ph.HE]} \BibitemShut {NoStop}%
\bibitem [{\citenamefont {Madau}\ and\ \citenamefont {Dickinson}(2014)}]{Madau_2014}%
  \BibitemOpen
  \bibfield  {author} {\bibinfo {author} {\bibfnamefont {P.}~\bibnamefont {Madau}}\ and\ \bibinfo {author} {\bibfnamefont {M.}~\bibnamefont {Dickinson}},\ }\bibfield  {title} {\bibinfo {title} {Cosmic star-formation history},\ }\href {https://doi.org/10.1146/annurev-astro-081811-125615} {\bibfield  {journal} {\bibinfo  {journal} {Annual Review of Astronomy and Astrophysics}\ }\textbf {\bibinfo {volume} {52}},\ \bibinfo {pages} {415} (\bibinfo {year} {2014})}\BibitemShut {NoStop}%
\bibitem [{\citenamefont {Abbott}\ \emph {et~al.}(2021{\natexlab{b}})\citenamefont {Abbott} \emph {et~al.}}]{LIGOScientific:2020kqk}%
  \BibitemOpen
  \bibfield  {author} {\bibinfo {author} {\bibfnamefont {R.}~\bibnamefont {Abbott}} \emph {et~al.} (\bibinfo {collaboration} {LIGO Scientific, Virgo}),\ }\bibfield  {title} {\bibinfo {title} {{Population Properties of Compact Objects from the Second LIGO-Virgo Gravitational-Wave Transient Catalog}},\ }\href {https://doi.org/10.3847/2041-8213/abe949} {\bibfield  {journal} {\bibinfo  {journal} {Astrophys. J. Lett.}\ }\textbf {\bibinfo {volume} {913}},\ \bibinfo {pages} {L7} (\bibinfo {year} {2021}{\natexlab{b}})},\ \Eprint {https://arxiv.org/abs/2010.14533} {arXiv:2010.14533 [astro-ph.HE]} \BibitemShut {NoStop}%
\bibitem [{\citenamefont {Cutler}\ and\ \citenamefont {Flanagan}(1994)}]{Cutler_1994}%
  \BibitemOpen
  \bibfield  {author} {\bibinfo {author} {\bibfnamefont {C.}~\bibnamefont {Cutler}}\ and\ \bibinfo {author} {\bibfnamefont {{\'{E} }.~E.}\ \bibnamefont {Flanagan}},\ }\bibfield  {title} {\bibinfo {title} {Gravitational waves from merging compact binaries: How accurately can one extract the binary's parameters from the inspiral waveform?},\ }\href {https://doi.org/10.1103/physrevd.49.2658} {\bibfield  {journal} {\bibinfo  {journal} {Physical Review D}\ }\textbf {\bibinfo {volume} {49}},\ \bibinfo {pages} {2658} (\bibinfo {year} {1994})}\BibitemShut {NoStop}%
\bibitem [{\citenamefont {Salafia}\ \emph {et~al.}(2015)\citenamefont {Salafia}, \citenamefont {Ghisellini}, \citenamefont {Pescalli}, \citenamefont {Ghirlanda},\ and\ \citenamefont {Nappo}}]{Salafia_2015}%
  \BibitemOpen
  \bibfield  {author} {\bibinfo {author} {\bibfnamefont {O.~S.}\ \bibnamefont {Salafia}}, \bibinfo {author} {\bibfnamefont {G.}~\bibnamefont {Ghisellini}}, \bibinfo {author} {\bibfnamefont {A.}~\bibnamefont {Pescalli}}, \bibinfo {author} {\bibfnamefont {G.}~\bibnamefont {Ghirlanda}},\ and\ \bibinfo {author} {\bibfnamefont {F.}~\bibnamefont {Nappo}},\ }\bibfield  {title} {\bibinfo {title} {Structure of gamma-ray burst jets: intrinsic versus apparent properties},\ }\href {https://doi.org/10.1093/mnras/stv766} {\bibfield  {journal} {\bibinfo  {journal} {Monthly Notices of the Royal Astronomical Society}\ }\textbf {\bibinfo {volume} {450}},\ \bibinfo {pages} {3549} (\bibinfo {year} {2015})}\BibitemShut {NoStop}%
\bibitem [{\citenamefont {Ghirlanda}\ \emph {et~al.}(2016)\citenamefont {Ghirlanda} \emph {et~al.}}]{Ghirlanda_2016}%
  \BibitemOpen
  \bibfield  {author} {\bibinfo {author} {\bibfnamefont {G.}~\bibnamefont {Ghirlanda}} \emph {et~al.},\ }\bibfield  {title} {\bibinfo {title} {{Short gamma-ray bursts at the dawn of the gravitational wave era}},\ }\href {https://doi.org/10.1051/0004-6361/201628993} {\bibfield  {journal} {\bibinfo  {journal} {Astron. Astrophys.}\ }\textbf {\bibinfo {volume} {594}},\ \bibinfo {pages} {A84} (\bibinfo {year} {2016})},\ \Eprint {https://arxiv.org/abs/1607.07875} {arXiv:1607.07875 [astro-ph.HE]} \BibitemShut {NoStop}%
\bibitem [{\citenamefont {Fong}\ and\ \citenamefont {Berger}(2013)}]{Fong_2013}%
  \BibitemOpen
  \bibfield  {author} {\bibinfo {author} {\bibfnamefont {W.}~\bibnamefont {Fong}}\ and\ \bibinfo {author} {\bibfnamefont {E.}~\bibnamefont {Berger}},\ }\bibfield  {title} {\bibinfo {title} {The locations of short gamma-ray bursts as evidence for compact object binary progenitors},\ }\href {https://doi.org/10.1088/0004-637X/776/1/18} {\bibfield  {journal} {\bibinfo  {journal} {The Astrophysical Journal}\ }\textbf {\bibinfo {volume} {776}},\ \bibinfo {pages} {18} (\bibinfo {year} {2013})}\BibitemShut {NoStop}%
\bibitem [{\citenamefont {Francisco J.~Virgili}\ and\ \citenamefont {Troja}(2011)}]{Virgili2011}%
  \BibitemOpen
  \bibfield  {author} {\bibinfo {author} {\bibfnamefont {P.~O.}\ \bibnamefont {Francisco J.~Virgili}, \bibfnamefont {Bing~Zhang}}\ and\ \bibinfo {author} {\bibfnamefont {E.}~\bibnamefont {Troja}},\ }\bibfield  {title} {\bibinfo {title} {Are all short–hard gamma-ray bursts produced from mergers of compact stellar objects?},\ }\href {https://doi.org/10.1088/0004-637X/727/2/109} {\bibfield  {journal} {\bibinfo  {journal} {The Astrophysical Journal}\ }\textbf {\bibinfo {volume} {727}},\ \bibinfo {pages} {109} (\bibinfo {year} {2011})},\ \bibinfo {note} {published 2011-01-11}\BibitemShut {NoStop}%
\bibitem [{\citenamefont {Thompson}\ and\ \citenamefont {Wilson-Hodge}(2022)}]{Thompson_2022}%
  \BibitemOpen
  \bibfield  {author} {\bibinfo {author} {\bibfnamefont {D.~J.}\ \bibnamefont {Thompson}}\ and\ \bibinfo {author} {\bibfnamefont {C.~A.}\ \bibnamefont {Wilson-Hodge}},\ }\bibfield  {title} {\bibinfo {title} {Fermi gamma-ray space telescope},\ }in\ \href {https://doi.org/10.1007/978-981-16-4544-0_58-1} {\emph {\bibinfo {booktitle} {Handbook of X-ray and Gamma-ray Astrophysics}}}\ (\bibinfo  {publisher} {Springer Nature Singapore},\ \bibinfo {year} {2022})\ pp.\ \bibinfo {pages} {1--31}\BibitemShut {NoStop}%
\bibitem [{\citenamefont {Zhang}(2019)}]{Zhang_2019}%
  \BibitemOpen
  \bibfield  {author} {\bibinfo {author} {\bibfnamefont {B.}~\bibnamefont {Zhang}},\ }\bibfield  {title} {\bibinfo {title} {The delay time of gravitational wave {\textemdash} gamma-ray burst associations},\ }\bibfield  {journal} {\bibinfo  {journal} {Frontiers of Physics}\ }\textbf {\bibinfo {volume} {14}},\ \href {https://doi.org/10.1007/s11467-019-0913-4} {10.1007/s11467-019-0913-4} (\bibinfo {year} {2019})\BibitemShut {NoStop}%
\bibitem [{\citenamefont {Gair}\ \emph {et~al.}(2023{\natexlab{b}})\citenamefont {Gair}, \citenamefont {Ghosh}, \citenamefont {Gray}, \citenamefont {Holz}, \citenamefont {Mastrogiovanni}, \citenamefont {Mukherjee}, \citenamefont {Palmese}, \citenamefont {Tamanini}, \citenamefont {Baker}, \citenamefont {Beirnaert} \emph {et~al.}}]{Gair_2023}%
  \BibitemOpen
  \bibfield  {author} {\bibinfo {author} {\bibfnamefont {J.~R.}\ \bibnamefont {Gair}}, \bibinfo {author} {\bibfnamefont {A.}~\bibnamefont {Ghosh}}, \bibinfo {author} {\bibfnamefont {R.}~\bibnamefont {Gray}}, \bibinfo {author} {\bibfnamefont {D.~E.}\ \bibnamefont {Holz}}, \bibinfo {author} {\bibfnamefont {S.}~\bibnamefont {Mastrogiovanni}}, \bibinfo {author} {\bibfnamefont {S.}~\bibnamefont {Mukherjee}}, \bibinfo {author} {\bibfnamefont {A.}~\bibnamefont {Palmese}}, \bibinfo {author} {\bibfnamefont {N.}~\bibnamefont {Tamanini}}, \bibinfo {author} {\bibfnamefont {T.}~\bibnamefont {Baker}}, \bibinfo {author} {\bibfnamefont {F.}~\bibnamefont {Beirnaert}}, \emph {et~al.},\ }\bibfield  {title} {\bibinfo {title} {The hitchhiker’s guide to the galaxy catalog approach for dark siren gravitational-wave cosmology},\ }\href {https://doi.org/10.3847/1538-3881/acca78} {\bibfield  {journal} {\bibinfo  {journal} {The Astronomical Journal}\ }\textbf {\bibinfo {volume} {166}},\ \bibinfo {pages} {22} (\bibinfo {year}
  {2023}{\natexlab{b}})}\BibitemShut {NoStop}%
\bibitem [{\citenamefont {Fishbach}\ \emph {et~al.}(2018)\citenamefont {Fishbach}, \citenamefont {Holz},\ and\ \citenamefont {Farr}}]{Fishbach:2018edt}%
  \BibitemOpen
  \bibfield  {author} {\bibinfo {author} {\bibfnamefont {M.}~\bibnamefont {Fishbach}}, \bibinfo {author} {\bibfnamefont {D.~E.}\ \bibnamefont {Holz}},\ and\ \bibinfo {author} {\bibfnamefont {W.~M.}\ \bibnamefont {Farr}},\ }\bibfield  {title} {\bibinfo {title} {{Does the Black Hole Merger Rate Evolve with Redshift?}},\ }\href {https://doi.org/10.3847/2041-8213/aad800} {\bibfield  {journal} {\bibinfo  {journal} {Astrophys. J. Lett.}\ }\textbf {\bibinfo {volume} {863}},\ \bibinfo {pages} {L41} (\bibinfo {year} {2018})},\ \Eprint {https://arxiv.org/abs/1805.10270} {arXiv:1805.10270 [astro-ph.HE]} \BibitemShut {NoStop}%
\bibitem [{\citenamefont {{Ronchini}}\ \emph {et~al.}(2022)\citenamefont {{Ronchini}}, \citenamefont {{Branchesi}}, \citenamefont {{Oganesyan}}, \citenamefont {{Banerjee}}, \citenamefont {{Dupletsa}}, \citenamefont {{Ghirlanda}}, \citenamefont {{Harms}}, \citenamefont {{Mapelli}},\ and\ \citenamefont {{Santoliquido}}}]{2022A&A...665A..97R}%
  \BibitemOpen
  \bibfield  {author} {\bibinfo {author} {\bibfnamefont {S.}~\bibnamefont {{Ronchini}}}, \bibinfo {author} {\bibfnamefont {M.}~\bibnamefont {{Branchesi}}}, \bibinfo {author} {\bibfnamefont {G.}~\bibnamefont {{Oganesyan}}}, \bibinfo {author} {\bibfnamefont {B.}~\bibnamefont {{Banerjee}}}, \bibinfo {author} {\bibfnamefont {U.}~\bibnamefont {{Dupletsa}}}, \bibinfo {author} {\bibfnamefont {G.}~\bibnamefont {{Ghirlanda}}}, \bibinfo {author} {\bibfnamefont {J.}~\bibnamefont {{Harms}}}, \bibinfo {author} {\bibfnamefont {M.}~\bibnamefont {{Mapelli}}},\ and\ \bibinfo {author} {\bibfnamefont {F.}~\bibnamefont {{Santoliquido}}},\ }\bibfield  {title} {\bibinfo {title} {{Perspectives for multimessenger astronomy with the next generation of gravitational-wave detectors and high-energy satellites}},\ }\href {https://doi.org/10.1051/0004-6361/202243705} {\bibfield  {journal} {\bibinfo  {journal} {A\&A}\ }\textbf {\bibinfo {volume} {665}},\ \bibinfo {eid} {A97} (\bibinfo {year} {2022})},\ \Eprint
  {https://arxiv.org/abs/2204.01746} {arXiv:2204.01746 [astro-ph.HE]} \BibitemShut {NoStop}%
\bibitem [{\citenamefont {{Foreman-Mackey}}\ \emph {et~al.}(2013)\citenamefont {{Foreman-Mackey}}, \citenamefont {{Hogg}}, \citenamefont {{Lang}},\ and\ \citenamefont {{Goodman}}}]{2013PASP..125..306F}%
  \BibitemOpen
  \bibfield  {author} {\bibinfo {author} {\bibfnamefont {D.}~\bibnamefont {{Foreman-Mackey}}}, \bibinfo {author} {\bibfnamefont {D.~W.}\ \bibnamefont {{Hogg}}}, \bibinfo {author} {\bibfnamefont {D.}~\bibnamefont {{Lang}}},\ and\ \bibinfo {author} {\bibfnamefont {J.}~\bibnamefont {{Goodman}}},\ }\bibfield  {title} {\bibinfo {title} {{emcee: The MCMC Hammer}},\ }\href {https://doi.org/10.1086/670067} {\bibfield  {journal} {\bibinfo  {journal} {PASP}\ }\textbf {\bibinfo {volume} {125}},\ \bibinfo {pages} {306} (\bibinfo {year} {2013})},\ \Eprint {https://arxiv.org/abs/1202.3665} {arXiv:1202.3665 [astro-ph.IM]} \BibitemShut {NoStop}%
\bibitem [{\citenamefont {Ashton}\ \emph {et~al.}(2019)\citenamefont {Ashton} \emph {et~al.}}]{bilby_paper}%
  \BibitemOpen
  \bibfield  {author} {\bibinfo {author} {\bibfnamefont {G.}~\bibnamefont {Ashton}} \emph {et~al.},\ }\bibfield  {title} {\bibinfo {title} {{BILBY: A user-friendly Bayesian inference library for gravitational-wave astronomy}},\ }\href {https://doi.org/10.3847/1538-4365/ab06fc} {\bibfield  {journal} {\bibinfo  {journal} {Astrophys. J. Suppl.}\ }\textbf {\bibinfo {volume} {241}},\ \bibinfo {pages} {27} (\bibinfo {year} {2019})},\ \Eprint {https://arxiv.org/abs/1811.02042} {arXiv:1811.02042 [astro-ph.IM]} \BibitemShut {NoStop}%
\bibitem [{\citenamefont {Romero-Shaw}\ \emph {et~al.}(2020)\citenamefont {Romero-Shaw} \emph {et~al.}}]{bilby_pipe_paper}%
  \BibitemOpen
  \bibfield  {author} {\bibinfo {author} {\bibfnamefont {I.~M.}\ \bibnamefont {Romero-Shaw}} \emph {et~al.},\ }\bibfield  {title} {\bibinfo {title} {{Bayesian inference for compact binary coalescences with bilby: validation and application to the first LIGO\textendash{}Virgo gravitational-wave transient catalogue}},\ }\href {https://doi.org/10.1093/mnras/staa2850} {\bibfield  {journal} {\bibinfo  {journal} {Mon. Not. Roy. Astron. Soc.}\ }\textbf {\bibinfo {volume} {499}},\ \bibinfo {pages} {3295} (\bibinfo {year} {2020})},\ \Eprint {https://arxiv.org/abs/2006.00714} {arXiv:2006.00714 [astro-ph.IM]} \BibitemShut {NoStop}%
\end{thebibliography}%

\end{document}